\documentclass[preprint,12pt,3p]{elsarticle}

\usepackage{amssymb}
\usepackage{amsmath}
\usepackage{comment}
\usepackage{xcolor}
\usepackage[numbers]{natbib}
\usepackage{tabularx}
\usepackage{tabularx,booktabs}
\usepackage{siunitx}    
\usepackage{ragged2e}
\usepackage{graphicx}
\usepackage{caption}
\usepackage{subcaption}
\usepackage[font=small,labelformat=parens,labelsep=space]{subcaption}
\usepackage{float}
\usepackage{cleveref}
\usepackage{tabularx, booktabs, array, ragged2e}
\journal{Physics Reports}

\begin{document}
\begin{frontmatter}

\title{Extreme nonlinear optics in optical fibers}

\author[label1,label2]{Mario Ferraro}
\author[label1]{Bertrand Kibler}
\author[label1]{Pierre Béjot}
\author[label3]{Frédéric Gérôme}
\author[label3]{Benoit Debord}
\author[label3]{Fetah Benabid}
\author[label4,label5]{Fabio Mangini}
\author[label5]{Stefan Wabnitz\corref{cor1}}
 \ead{stefan.wabnitz@uniroma1.it}
\cortext[cor1]{Corresponding author}

\affiliation[label1]
{organization={Laboratoire Interdisciplinaire Carnot de Bourgogne (ICB), UMR 6303},
             addressline={Université Bourgogne Europe, CNRS},
             city={Dijon},
             postcode={21000},
             country={France}}

 \affiliation[label2]{organization={Physics Department, University of Calabria},
             addressline={Via P. Bucci},
             city={Rende},
             postcode={87036},
             country={Italy}}

 \affiliation[label3]{organization={GPPMM Group, Xlim Research Institute, UMR 7252, University of Limoges, CNRS},
             addressline={123 Avenue Albert Thomas},
             city={Limoges},
             postcode={87060},
             country={France}}

 \affiliation[label4]{organization={Department of Engineering, Niccolò Cusano University,},
             addressline={via Don Carlo Gnocchi 3},
             city={Rome},
             postcode={00166},
             country={Italy}}

 \affiliation[label5]{organization={Dipartimento di Ingegneria dell’Informazione, Elettronica e Telecomunicazioni, Sapienza Universit`a di Roma},
             addressline={via Eudossiana 18},
             city={Rome},
             postcode={00184},
             country={Italy}}

\begin{abstract}
This paper reviews the field of extreme nonlinear optics in optical fibers, highlighting key phenomena and advancements. It discusses multiple ionization effects caused by femtosecond laser pulses that generate plasma and induce permanent material modifications, as well as plasma luminescence and its dependence on material imperfections. The formation and dynamics of plasma filaments, including helical structures, are explored, along with the rainbow spiral emission pattern useful in communications and particle manipulation.
The review covers the generation of spatial-temporal waves, supercontinuum broadening, and advanced modeling techniques, such as multimode unidirectional pulse propagation equations for describing optical pulse evolution. Experimental demonstrations involving discretized conical waves and supercontinuum generation optimization are detailed. The paper emphasizes the unique capabilities of photonic crystal fibers, especially hollow-core variants, in achieving broad supercontinua and Raman frequency combs, ultrashort pulse compression, high-harmonic generation, plasma formation, and nonclassical light production.
Our outlook highlights ongoing research into spatiotemporal helicon waves, ultrashort pulse propagation, vacuum ultraviolet and mid-infrared supercontinuum generation, and innovative fiber technologies. Future directions focus on enhancing fiber performance, understanding multimodal wave dynamics, and expanding applications in telecommunications, sensing, and quantum science.
\end{abstract}

\begin{keyword} Nonlinear optics \sep Optical fibers \sep Supercontinuum generation \sep Plasma generation \sep Raman scattering \sep Hollow-core fibers \sep Conical emission \sep Self-focusing



\end{keyword}

\end{frontmatter}


\section{Introduction}
\label{sec:intro}

The regime of extreme nonlinear optics occurs when intense laser fields drive matter far beyond the perturbative regime, leading to phenomena such as high-harmonic generation (HHG), multi-photon ionization (MPI), self-focusing, conical emission, and plasma ignition, to name a few.
Seminal studies on extreme nonlinear optical phenomena were carried out by Von der Linde and Schüler \cite{von1996breakdown} and by Li et al. \cite{li1999ultrafast}. These works established the breakdown thresholds and ultrafast electron dynamics in dielectrics under femtosecond excitation, providing a foundation for understanding plasma generation and nonlinear absorption in condensed media. Nowadays, extreme nonlinear optics is extensively studied across several optical platforms \cite{wegener2005extreme,chin2005propagation}, ranging from gases to bulk solids such as dielectrics \cite{stuart1996nanosecond,lenzner1998femtosecond, milchberg2014extreme}, condensed matter systems \cite{kolesik2004nonlinear}, and even natural gems \cite{filosa2023nonlinear}.

Among the several extreme nonlinear optical phenomena, conical emission is particularly spectacular: it produces a characteristic angular distribution of the emitted light, which appears as a rainbow \cite{nibbering1996conical,faccio2006conical}. The power needed for generating conical emission is so high that, in general, this effect is accompanied by the ionization of the material and the consequent ignition of a plasma \cite{keldysh1964ionization}. In particular, owing to a phenomenon known as self-channeling, the plasma assumes the shape of a filament \cite{CouaironMysy2007}. Interestingly, curved plasma structures were also generated by exploiting the non-diffractive properties of Airy beams, as first demonstrated by Polynkin et al. in 2009 \cite{Polynkin2009}. Such curved plasma structures have attracted significant interest for both atmospheric physics and photonics applications \cite{Kasparian2008}.

Another remarkable manifestation of extreme nonlinear optics involving frequency conversion is HHG, which was originally observed in gases \cite{ferray1988multiple,mcpherson1987studies} and later demonstrated in solids \cite{ghimire2019high}. In this process, an intense laser field drives electrons far from equilibrium, leading to the emission of coherent radiation at multiples of the driving frequency, enabling the generation of attosecond pulses and coherent light extending up to the X-ray region.

In recent years, there has been a growing interest in studying extreme nonlinear optical effects within guided systems such as optical fibers, where the interplay between light confinement and high-intensity interactions gives rise to unique and previously inaccessible regimes of nonlinear dynamics. The interplay between waveguiding and extreme nonlinearities enables the generation of complex spatiotemporal structures and ultrabroadband spectra, laying the groundwork for advanced applications in fiber-based supercontinuum generation, high-harmonic generation, and ultrafast photonics.

In this regard, hollow-core photonic crystal fibers (HCPCFs) offer a unique platform for extreme nonlinear optics, by guiding light in air rather than in solid glass, resulting in extremely low nonlinear absorption and high damage thresholds \cite{wang2013hollow, Gerome30}. When filled with gases or other nonlinear media, these fibers enable the observation and control of ultrafast phenomena such as Raman frequency combs \cite{Benabid3,Benabid5}, soliton dynamics \cite{Gerome19}, supercontinuum generation \cite{Gerome22,Benabid11}, and high-harmonic generation \cite{Benabid12}, while minimizing material-induced losses. The unique dispersion and modal properties of HCPCFs also allow for the generation of exotic spatiotemporal structures, including plasma filaments and dispersive wave emission \cite{Brahms:19}, with fine control over phase-matching conditions. Consequently, hollow-core fibers provide an ideal bridge between bulk (or gas) nonlinear optics and solid-core waveguides, expanding the possibilities for high-intensity light–matter interactions and applications in ultrafast photonics, attosecond science, and quantum light generation.

Extreme nonlinear optical effects have also been studied in standard optical fibers, whereby light is guided via the phenomenon of total internal reflection. In particular, several studies focused on multimode optical fibers, which are essential for bridging nonlinear optics in bulk media and single-mode fibers. Understanding the transition between the two fields is complicated by intermodal nonlinear processes and spatiotemporal couplings. Interestingly,  striking phenomena have recently been revealed in such waveguides. Some of these were previously observed in bulk media with ultrashort pulses, like multiphoton ionization effects \cite{Ferraro2022}, whereas others take place only thanks to the light-guiding properties, such as discrete conical emission \cite{kibler2021discretized,stefanska2023experimental}, or to the cylindrical geometry of the optical fibers, e.g., rainbow spiral emission \cite{Mangini2021-spiral}.

In this paper, we present a comprehensive overview of the advancements and key phenomena associated with extreme nonlinear optics in optical fibers, with a particular focus on multimode fibers and specialized fiber structures such as photonic crystal fibers (PCFs), especially their hollow-core variants. We discuss how intense laser pulses, particularly femtosecond pulses, induce multiple ionization effects within the fiber material, leading to plasma formation that permanently alters the glass matrix. These plasma effects can cause luminescence, with color variations linked to material defects, and are observed to follow periodic patterns in graded-index (GRIN) fibers due to self-imaging phenomena.

A significant part of the paper describes plasma filaments—localized regions of plasma that influence the fiber's refractive index, often causing defocusing effects that balance Kerr nonlinearity. Researchers have generated helical plasma filaments, offering potential for novel fiber structures and optical memory applications. The dynamics of these filaments depend on parameters such as the tilt angle of the input laser beam, which can influence their formation and behavior.

The phenomenon of rainbow spiral emission is highlighted, where injected light generates a supercontinuum spectrum that organizes into spiral patterns, with various colors distributed along the spiral. This process depends on initial injection conditions, and has promising applications in optical communications and in the manipulation of particles.

The exploration of spatial-temporal waves—complex wave structures that involve both spatial and temporal dimensions—is emphasized, especially their generation under high-intensity conditions near self-focusing thresholds. These waves can be designed as invariant wave packets through the superposition of modes in quadric spaces, and are significant in fundamental physics and quantum technologies.

On the modeling front, the paper discusses the use of multimode unidirectional pulse propagation equations (MM-UPPE), which simulate supercontinuum generation more accurately than earlier models by capturing complex nonlinear responses and dispersion effects. These models help analyze how the electric field evolves during propagation in multimode fibers, especially when a Gaussian pulse interacts with multiple orbital angular momentum (OAM) modes, resulting in spectral broadening and intense wave formations.

Experimental breakthroughs include the demonstration of discretized conical waves and their analysis through high-power femtosecond lasers, confirming theory with observations such as strong spectral amplification toward visible wavelengths. Optimization techniques for supercontinuum generation involve adjusting fiber taper geometries and initial mode coupling to maximize broadening, which in some cases extends into the mid-infrared spectrum at high powers.

Next, the paper emphasizes the exceptional nonlinear capabilities of HCPCFs. These fibers guide light predominantly in air, resulting in very low loss and high damage thresholds compared to solid-core silica fibers. When functionalized with a nonlinear gaseous medium, they enable the generation of extremely broad light sources spanning from vacuum ultraviolet to mid-infrared, involving Raman frequency combs or supercontinua driven by mechanisms such as soliton self-compression, soliton-ionization dynamics and dispersive wave emission. HCPCFs have also been instrumental in ultrashort pulse compression, achieving pulse durations close to single-cycle and attosecond regimes, which are vital for high-precision applications.

HHG is another crucial topic covered, with HCPCFs enabling harmonic orders extending into the extreme ultraviolet region. By using specific configurations, these fibers even enable the generation of unprecedentedly high flux photons, thereby facilitating attosecond science. Plasma generation in these fibers is also notable, as they can sustain stable plasmas via microwave discharges or ultrafast ionization processes without damaging the fiber structure, opening pathways for extreme nonlinear phenomena.

The generation of nonclassical light, including photon pairs and single photons, was demonstrated within noble gas-filled HCPCFs, emphasizing their relevance for quantum optics and quantum information science. These fibers support collective emission processes like superradiance, offering new opportunities for quantum light sources.

The outlook underscores future directions, such as understanding multimodal propagation, enhancing fiber fabrication techniques, and expanding applications in telecommunications, sensing, and quantum science. Emerging research on spatiotemporal helicon waves—complex wave phenomena with unique propagation characteristics—has been explored, presenting advanced possibilities in photonics. Studies on ultrashort pulse propagation and mid-infrared supercontinuum generation continue to push the boundaries of broadband light sources, with recent demonstrations of multi-octave supercontinuum generation in chalcogenide glass fibers. 

Beyond the study of plasma effects, MPI in multimode fibers is emerging as a pathway for direct material modification and structure writing within the fiber core. The interplay between femtosecond micromachining, plasma filament control, and ultraluminescence diagnostics points toward a new generation of smart fibers, where light itself becomes an active agent for fabrication, sensing, and manipulation.

Finally, ongoing developments in hollow-core fiber technologies aim to improve light guidance, support soliton formation, and enable ultrashort pulse compression and harmonic generation at new wavelengths. Collectively, these efforts aim to establish new paradigms in nonlinear fiber optics, with broad implications across scientific and technological domains.

This work is structured as follow: in Sec. \ref{sec:rome}, we present some recent experimental results of MPI effects in standard solid core multimode fibers (MMFs); then, Sec. \ref{sec:dijon} focuses on the topic of space-time wavepackets and frequency conversion around the self-focusing threshold in (solid-core) MMFs; whereas in Sec. \ref{sec:limoges}, we discuss on nonlinear optics in gas-filled HCPCFs. Finally, Sec. \ref{sec:conclusion} is dedicated to conclusions and perspectives.


\section{Multiphoton ionization effects in multimode fibers}
\label{sec:rome}

In this section we discuss MPI effects in standard multimode fibers that are observed by using intense femtosecond infrared (IR) or visible laser pulses. Generally speaking, MPI in glasses arises from the nonlinear absorption of multiple photons, whose combined energy exceeds the bandgap (usually in the ultraviolet (UV) range), thus generating seed electrons in the conduction band. Subsequent carrier multiplication via the mechanism of avalanche ionization drives the system to reach the critical plasma density, at which point a localized electron–hole plasma is established. Note that, unlike avalanche multiplication in avalanche photodiodes, which requires a sustained DC electric field, avalanche ionization during femtosecond laser interaction with glass is driven by the intense oscillating laser field \cite{stuart1996nanosecond}. The optical response of the plasma is often described within the Drude model, which relates the free-electron density and collision frequency to transient changes in absorption and refractive index \cite{schmitz2012full}. 

One of the most relevant differences between MPI in gases and in solid platforms, such as standard optical fibers, is that in the former case, the medium generally recovers its initial state once the laser source is switched off. Whereas, in the latter, plasma generation leads to irreversible modifications of the material. Specifically, with standard fibers, plasma can induce permanent modifications of the glass matrix, including densification, defect formation, and refractive index changes, depending on the irradiation regime. The deterioration of the fiber material is the first topic addressed in this section. Then, we focus on the features of plasma and, in particular, its luminescence and the formation of curved plasma filaments, obtained by exploiting the cylindrical geometry of the fiber \cite{Mangini2022-helical}. Finally, we present the intriguing phenomenon of rainbow spiral emission \cite{Mangini2021-spiral}.

\subsection{Deterioration of the fiber material}
\label{sec:damages}

MPI is widely exploited in ultrafast laser micromachining of glasses, since it enables energy deposition in a highly localized volume without affecting the surrounding material. This mechanism has provided a pathway to induce controlled refractive index changes and is, to date, of widespread use. Compared to glass material modifications using UV light, MPI permits reducing the photon energy which is needed for material structure modifications, thus reducing their size, and improving the precision of laser-based micromachining and ablation. This precision is of particular relevance in the development of photonic components, where multiphoton processes are employed to inscribe optical waveguides and tailor the refractive index profile of optical fibers, e.g., to directly write fiber Bragg gratings with high spatial resolution.

However, when dealing with high-peak-power IR lasers, owing to the presence of thermal effects, one may risk not being able to control the features of the induced damages, e.g., their size or rate of formation. Although this aspect is has a significant impact for the machining of plastic optical fibers \cite{kiedrowski2024comparing}, it might become quite relevant even for standard silica fibers, owing to photon-phonon coupling effects \cite{rottwitt2005analyzing}. 
In order to overcome this issue, the use of femtosecond laser pulses has been proposed. Ultrafast laser irradiation minimizes detrimental thermal effects in optical fibers, since the lattice (phonon) response occurs on much longer timescales than the femtosecond laser pulse duration.


The most common technique for optical fiber micromachining consists of injecting a laser beam orthogonally to the fiber axis and focusing it into the fiber core or cladding. By properly switching on and off the laser and moving the fiber by using a stage, one may craft a microstructure at will within the fiber volume. For reviews of glass and optical fiber micromachining, the reader may refer to Ref. \cite{ran2021fiber,pallares2020optical,he2021review}.

\begin{figure}
\begin{center}
\includegraphics[width=0.95\columnwidth]{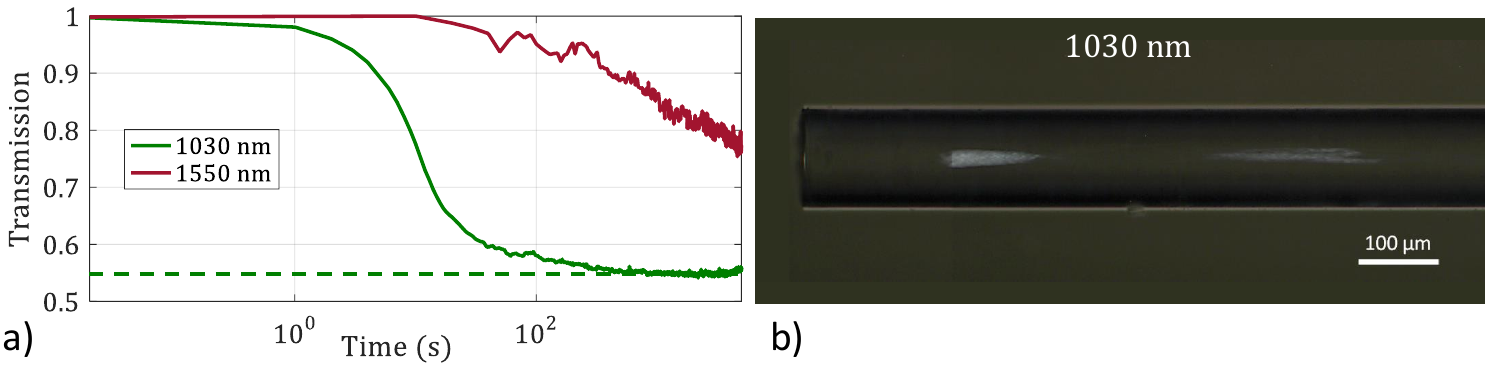}
\caption{Deterioration of the fiber material because of MPI. a) Transmission quenching vs. time. The dashed horizontal line represents the steady value of transmission reached, once that the deterioration process is over. b) Optical microscope image of the damage formed in the vicinity of the input GRIN fiber tip. The white bar corresponds to 100 $\mu$m. Details are available in Ref. \cite{Ferraro2022}.}
\label{fig:damages}
\end{center}
\end{figure}

A different configuration was studied in a recent paper, whereby the laser beam was injected parallel to the fiber axis \cite{Ferraro2022}. It was found that the nonlinear beam dynamics undergoes a complex evolution on a time scale of several hours. Such an evolution is irreversible, since it is due to plasma-driven material deterioration. The long timescale evolution of the beam dynamics depends on several factors, e.g., the regime of chromatic dispersion. However, in all cases the occurrence of progressive material deterioration leads to a drop in the transmission from the optical fiber. This is shown in Fig. \ref{fig:damages}a in the case of a standard graded-index (GRIN) MMF. Here two irradiation wavelengths were used, namely 1030 and 1550 nm, corresponding to the normal and anomalous dispersion regime, respectively. One may notice that the transmission drop at 1030 nm is much faster than that at 1550 nm. This is likely due to the fact that, at shorter wavelengths, a lower number of photons is needed to trigger MPI processes, the silica bandgap being in the UV range.
The transmission drop eventually stops at a steady value, indicated by a dashed line in Fig. \ref{fig:damages}a.

A notable property that was found in \cite{Ferraro2022} is that the material modification induced at the self-focusing point is spontaneously replicated along the GRIN fiber axis, with a periodicity of the order of hundreds of microns. An example is shown in Fig. \ref{fig:damages}b. Note that only a few replicas are observed. This is because the beam progressively loses its power upon propagation, down to the point where plasma can no longer be ignited.
The periodicity of the structure turns out to be provided by the phenomenon of spatial self-imaging, the spatial equivalent of the temporal Talbot effect \cite{Hansson2020,Karlsson1992}. In short, the parabolic profile of the core refractive index of GRIN fibers makes any light beam self-replicating over a period ($\Gamma$), which only depends on the optical and geometrical characteristics of the fiber, i.e.,
\begin{equation}
    \Gamma = 2\pi/\sqrt{g},
    \label{eq:gamma}
\end{equation}
where $g=2\Delta/r_c^2$ is the index grading parameter, being $\Delta$ and $r_c$ the core-cladding refractive index difference and core radius, respectively. 

\subsection{Plasma luminescence}
\label{sec:luminescence}

\begin{figure}
\begin{center}
\includegraphics[width=0.65\columnwidth]{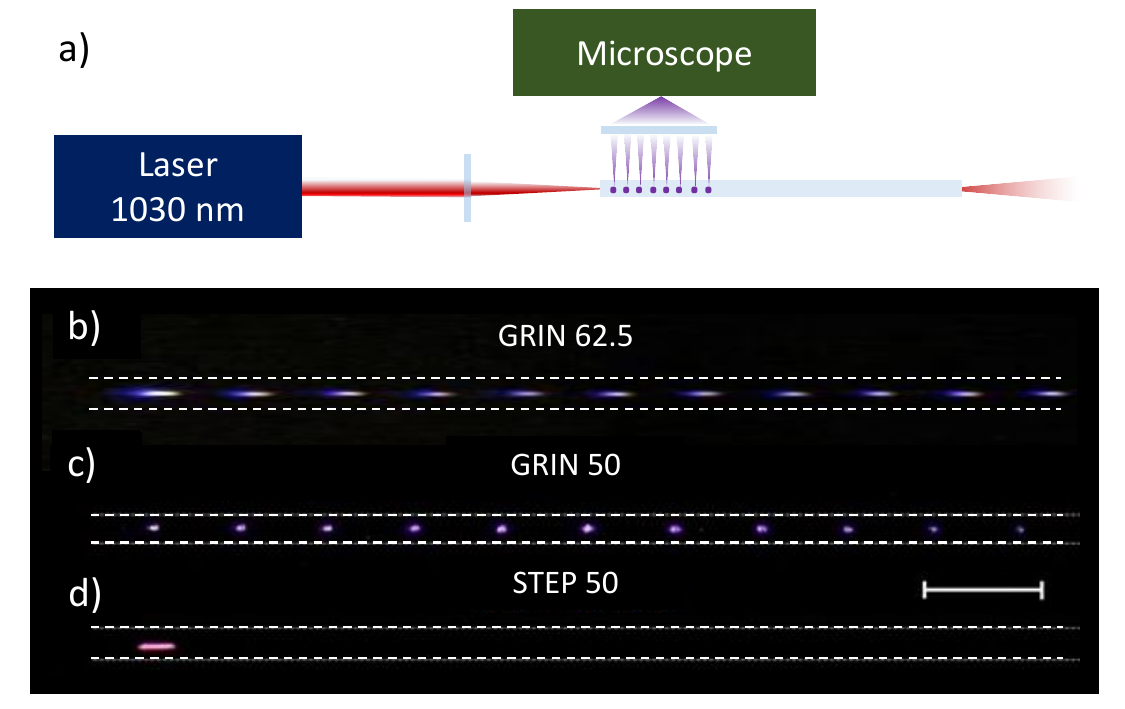}
\caption{a) A sketch of the experimental setup used for generating UL in silica fibers using 5-photon absorption \cite{Mangini2020}. b-d) Digital microscope images of the luminescence scattered from GRIN with 62.5 $\mu$m (b) and 50.0 $\mu$m (c), core diameter, and 50.0 $\mu$m step-index fibers (d). The dashed lines denote the cladding-air interface, while the scale bar corresponds to 1 mm. Further details are available in Ref. \cite{Mangini2020}.}
\label{fig:luminescence}
\end{center}
\end{figure}

One of the most well-known properties of plasma is provided by its luminescence. For instance, in astrophysics, plasma in stars emits characteristic colors that depend on star temperature and composition, giving rise to the observed spectral classes from red to blue. Similarly, plasma generated via MPI in optical fibers appears as bright visible light, which is detectable by the naked eye. Indeed, such plasma light emission results from a process of up-conversion luminescence (UL) of atoms and molecules which form the fiber material. In a recent work \cite{Ferraro2022}, the plasma generated in multimode silica fibers was imaged with a microscope by using the setup sketched in Fig. \ref{fig:luminescence}a. However, it should be mentioned that the phenomenon of UL does not necessarily require the ionization of the fiber material. In fact, some material defects are luminescent per se \cite{Girard2019}. The visualization of UL excited via multi-photon absorption (MPA) was reported in \cite{Mangini2020} in the case of silica, and in \cite{ferraro2024observation} in the case of non-silica fibers. In short, the "color" of plasma is inherited by that of the UL of the material defects. The spectrum of the UL depends on the material, just like the color of plasma depends on the nature of the gas which is ionized. As such, one may have UL light that spans from the red to the violet. A nice example is provided the emission of green luminescence in Ba-doped fluorosilicate fibers \cite{pietros2023investigation}.

Particularly spectacular is the case of GRIN fibers, where a spatially periodic pattern of UL is formed along the fiber axis. This is shown in Fig. \ref{fig:luminescence}b,c in the case of standard GRIN 62.5/125 and 50/125 fibers, respectively. Analogously to Fig. \ref{fig:damages}b, the periodic structure of the UL is due to the parabolic shape of the refractive index profile of GRIN fibers, that gives rise to the spatial self-imaging effect.
The difference in the periodicity of the UL pattern in Fig. \ref{fig:luminescence}b and c is due to the dependence of $\Gamma$ on $g$, see eq. (\ref{eq:gamma}). Indeed, the two fibers shown there have different core sizes and numerical apertures (i.e., $\Delta$). As far as the difference in the shape of the UL pattern is concerned, this is likely due to chromatic aberrations introduced by the microscope objective or by the fibers themselves. Regardless, the measurement of $\Gamma$ in GRIN fibers using the simple microscopic imaging setup in Fig. \ref{fig:luminescence}a permits the estimation of the value of the refractive index difference $\Delta$ via eq. (\ref{eq:gamma}). Incidentally, it is worth mentioning that the spatial self-imaging effect trivially occurs in two-mode fibers as well. By exploiting the measurement of $\Gamma$ in that case, one may estimate the cutoff frequency for single mode propagation \cite{Mangini2021}.

For the case of GRIN silica fibers as reported in Fig. \ref{fig:luminescence}b, a watchful eye may notice that the color of luminescence is not uniform. Indeed, the color of UL consists of three contributions: violet, cyan, and red, respectively. The first two are due to defects ascribed to the presence of the Ge doping, which gives the parabolic profile to the fiber core; the latter is due to an intrinsic defect of the silica lattice structure. Indeed, in the case of step-index fibers with pure silica core, the UL appears as bright red light (see Fig. \ref{fig:luminescence}d). The absence of Ge doping leads, on the one hand, to the monochromaticity of the UL and, on the other hand, to the absence of the self-imaging effect (and thus to the spatial periodicity of UL).

\subsection{Plasma filaments}
\label{sec:helical}

In silica, the presence of plasma reduces the material refractive index. This introduces a defocusing term in the optical beam propagation equation or, equivalently, a negative lens. To the contrary, Kerr nonlinearity tends to self-focus beams that propagate in the medium, which is akin to a positive lens, whose focusing efficiency increases with optical power. In standard optical conditions, the nonlinear contribution to the refractive index is rather weak. When it comes to very high powers, though, that is, in the extreme nonlinear optics regime, the nonlinear term becomes so relevant that it can reach the same order of magnitude as that of the linear refractive index difference between core and cladding. It may happen, for instance, that the Kerr term exactly balances the defocusing plasma term. When this happens, the beam may propagate without varying its shape. In short, light propagation resembles that of a liquid in a pipe or channel, which can propagate over distances longer than several thousands of wavelengths, without undergoing any diffractive spreading. For this reason, the balance between Kerr nonlinearity and plasma defocusing is called the self-channeling effect~\cite{Korn1995}. In this situation, the light beam is surrounded by plasma and, for this reason, self-channeling is often dubbed plasma filamentation. Here, we do not aim to go into the details of this effect, which is masterly described in Refs. \cite{CouaironMysy2007,Berge2007}. Instead, we limit ourselves to reporting on its observations in standard silica fibers.

\begin{figure}
\begin{center}
\includegraphics[width=0.85\columnwidth]{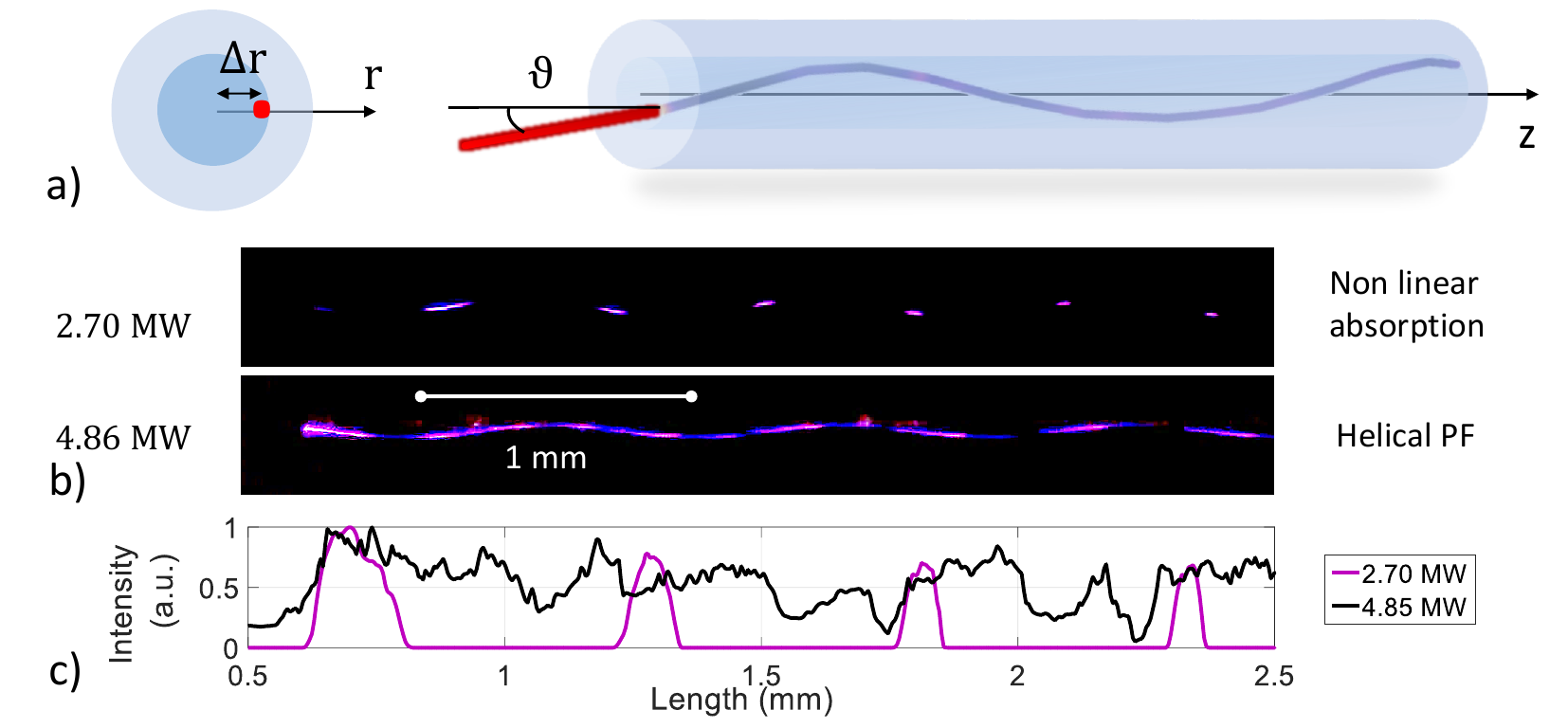}
\caption{Helical plasma filaments in optical fibers. a) Geometry of beam injection and propagation. b) Comparison between the UL generated at power below self-channeling (top) and the helical plasma filament (bottom). The values of the left indicate the beam peak power. c) Intensity profile of the scattered light, obtained as the integral of corresponding images in b). For additional information see Ref. \cite{Mangini2022-helical}.}
\label{fig:helix}
\end{center}
\end{figure}

The first reports of plasma filaments in standard optical fibers date back to Cho et al. in 1998 \cite{cho1998observation}. Later on, the same research group proposed to exploit the material modifications induced by plasma filaments to develop optical memories \cite{cho1999fabrication,cho2002situ,cho2003fabrication}. 
More recently, experimental and numerical studies have demonstrated the generation of \textit{helical} plasma filaments within standard multimode optical fibers \cite{Mangini2022-helical}. At variance with former works that only involved straight plasma filaments, in these studies the formation of the helical plasma structures arises from the interplay between conventional skew-rays propagation and the self-channeling phenomenon. The key idea of the method is the injection of intense laser pulses into the fiber with both a lateral offset and an angular tilt relative to the fiber axis.  Specifically, the beam must be injected inside the fiber core in the vicinity of the cladding interface, with an angle that favors the propagation of a skew-ray protected by the different topology between core and cladding \cite{HASHIMOTO1982,Sugon2015,Webster2017}. In this way, the light field is forced to travel in the core, close to the core–cladding boundary and in a helical path, rather than along the central axis, and so does the plasma filament generated by self-channeling (see Fig. \ref{fig:helix}a)~\cite{Bliokh2008}. This happens in the first millimeters of propagation; at longer distances, the optical power becomes too weak that neither MPI nor beam self-channeling may take place.

Interestingly, in step-index fibers the helix pitch can be readily controlled by adjusting the input tilt angle, like for conventional skew-rays. Conversely, in GRIN fibers the spatial periodicity of this helix is inherently determined by the self-imaging period (cfr. the images in Fig. \ref{fig:helix}b,c). 
In practical terms, the controlled inscription of helical plasma filaments opens the door to the fabrication of novel fiber structures-such as twisted or multi-core configurations ~\cite{alexeyev2013helical,beravat2016twist,russell2017helically} - alongside the development of volumetric optical storage systems and integrated active elements for high-performance fiber lasers. 
As mentioned above, the generation of plasma within optical materials leads to permanent and irreversible modifications of the refractive index, thus enabling high-precision microstructuring at the micrometer scale. Specifically, the capability of igniting helical plasma filaments may be advantageous for the fabrication of complex fiber architectures, including helical-shaped photonic crystal fibers and three-dimensional optical memory devices.

\subsection{Rainbow spiral emission}
\label{sec:rainbow}

\begin{figure}[!h]
\begin{center}
\includegraphics[width=0.85\columnwidth]{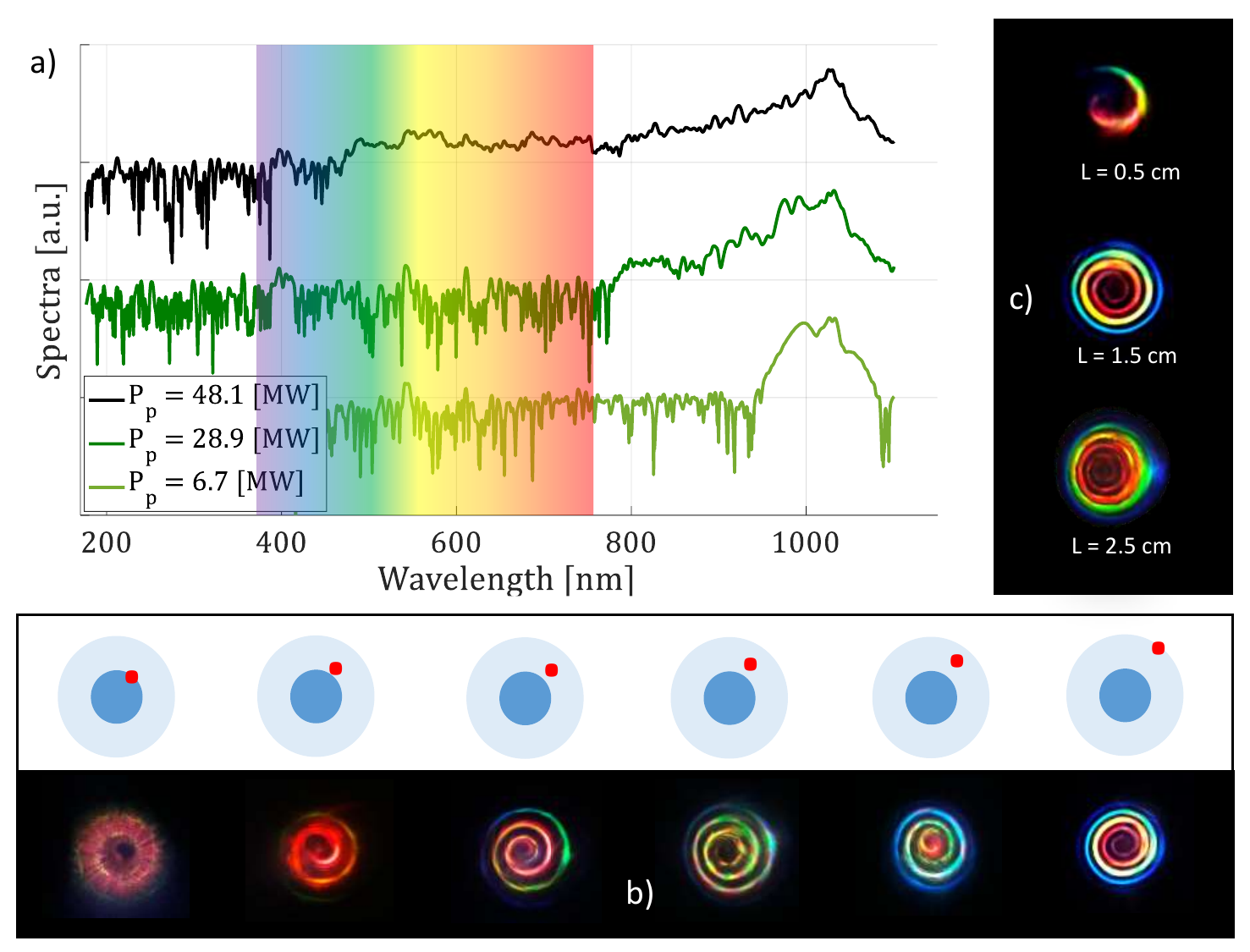}
\caption{Rainbow spiral emission. a) SCG spectra at different input peak powers. b) Dependence of the far-field patterns (bottom row) on the input beam position (red dot in the upper row). c) Rainbow spiral emission at different fiber lengths. Additional information are available in Ref. \cite{Mangini2021-spiral}.}
\label{fig:Rainbow}
\end{center}
\end{figure}

So far, we have discussed the properties of UL that is scattered out  from the fiber sides at the occurrence of MPI. The color of such light is that of the material luminescent defects. However, during its propagation in the MPI regime, a near-IR beam undergoes several nonlinear effects that may modify its spectrum in a non-trivial way. In particular, the high powers used to achieve MPI give rise to the phenomenon of supercontinuum generation (SCG). In short, the input narrow spectral profile of the beam quickly broadens, up to covering the whole visible spectral range. This is shown in Fig. \ref{fig:Rainbow}a in the specific case where light is injected in the cladding of a standard GRIN fiber. 
Such an unusual input condition is rather interesting, because light tends to remain confined within the cladding for distances of a few centimeters, i.e., about one order of magnitude longer than the survival of plasma filaments. As a consequence, light is somehow forced to spin around the fiber core, acquiring a \emph{transverse} orbital angular momentum. Such a light spinning results in the spontaneous formation of an Archimedean spiral pattern in the far-field at the output of the fiber. This is a linear effect provided by the conservation of the orbital angular momentum and the refractive index mismatch between glass and air \cite{Mangini2021-spiral}. 

As a result, the plasma filament which is formed in the MPI regime is also helicoidal, i.e., it is nearly identical to that shown in Fig. \ref{fig:helix}. In addition, the spiral shape of the output far-field merges with SCG, giving rise to the multicolor spiral patterns which are shown in Fig. \ref{fig:Rainbow}b. As can be seen, the distribution of colors of the spectral components of SC light along the spiral curve depends on the injection conditions. In particular, when the beam is injected on the cladding-air interface, the colors spontaneously organize, giving rise to a spiral-shaped rainbow (last panel on the right in Fig. \ref{fig:Rainbow}b). This phenomenon was dubbed rainbow spiral emission in Ref. \cite{Mangini2021-spiral}. Besides his beauty, the generation of a rainbow spiral beam may have implications for various applications in communications and optical manipulation of particles or tweezers, which are based on light-carrying orbital angular momentum. 

In this regard, it is worth mentioning that the self-organization of colors in a rainbow fashion turns out to be nearly independent of the laser wavelength and polarization. In addition, it was shown that the length of the spiral arm increases as the fiber length grows larger (see Fig. \ref{fig:Rainbow}c). Specifically, within the experimental conditions used in Ref. \cite{Mangini2021-spiral}, spiral emission was sustained for fiber lengths up to approximately 5 cm; beyond this distance, the internal phase distribution of the beam becomes distorted and it no longer supports spiral emission.

Besides rainbow spiral emission, SCG is a phenomenon that takes place in optical powers even at power levels way below the self-focusing and the MPI threshold. For a review of SCG in optical fibers see Ref. \cite{dudley2010supercontinuum}.
In the extreme nonlinear optics regime, SCG and, in general, frequency conversion effects are very interesting but yet relatively unexplored. In the next section, we will go through some recent advances in this field. In particular, at variance with the rainbow spiral emission, we will consider the case of light propagating in the fiber core, whereby the beam dynamics merges spectral (or temporal) and spatial degrees of freedom.

\section{Space-time wavepackets and ultrabroadband frequency conversion around the self-focusing threshold in multimode optical fibers}
\label{sec:dijon}



Structured light has found a wide variety of applications in many fields, including fundamental physics, telecommunications, particle manipulation, quantum information technology, laser micromachining, and high-resolution imaging \cite{zhan2009cylindrical,dholakia2011shaping,rubinsztein2016roadmap,forbes2021structured,he2022towards,shen2023roadmap}. Propagation-invariant or non-diffracting beams are an important class of structured optical fields that have received considerable attention. These beams are special solutions to Maxwell’s equations that maintain their shape during propagation \cite{durnin1987diffraction}. The simplest and best-known (theoretically) diffraction-free beams in bulk media are monochromatic plane waves. In addition to this trivial solution, monochromatic invariant beams have been extensively studied in free-space, including Bessel beams, Mathieu beams, self-accelerating Airy beams, and others. Beams carrying OAM -i.e., vortex beams- are also solutions to Maxwell’s equations and have attracted growing interest in recent years. They are characterized by a phase singularity at the center, a helical wavefront, and their topological charge (or winding number) $\ell$, which is an integer that counts the number of rotations of the wavefront over one propagation wavelength. While not all such beams are diffraction-free, some are (e.g., higher-order Bessel beams). Propagation-invariant pulsed beams have also been extensively studied. These solutions are not only diffraction-free, as their monochromatic counterparts, but are also insensitive to dispersion (e.g., the Bessel-X pulses in bulk media). These fully localized waves are non-dispersive and non-diffractive, owing to their intrinsically coupled spatio-temporal properties, which cancel the effects of spatial diffraction and temporal dispersion \cite{yessenov2022space,bejot2022quadrics}. These wavepackets exhibit unique characteristics, including controllable group velocities in free-space and unusual refractive phenomena. Thanks to recent innovations in spatiotemporal Fourier synthesis, corresponding experimental developments have emerged. These wavepackets can be generated linearly by shaping their spatio-temporal spectrum \cite{cruz2022synthesis,su2025space}. They can also emerge spontaneously when a sufficiently intense short pulse propagates non-linearly in a bulk medium or multimode waveguide, close to the self-focusing threshold of the material \cite{conti2003nonlinear,kibler2021discretized}.

\subsection{Invariant space-time linear wavepackets: Definition, construction and properties}
\label{sec:quadrics}

In this section, we review the general approach to structuring three-dimensional ultrashort light pulses in dispersive waveguiding structures such as MMFs, as recently developed in Ref. \cite{bejot2022quadrics}. Specifically, we illustrate with spatiotemporal (ST) wavepackets that are invariant under uniform rectilinear motion and spatial rotation (also known as ST helicon wavepackets \cite{bejot2021spatiotemporal}). These wavepackets can be constructed from the coherent superposition of modes lying on quadric surfaces in the frequency-resolved modal space. The nature of these surfaces depends on the dispersion regime considered and the group velocity of the wavepacket. Geometrically studying the quadrics allows to construct invariant ST wavepackets that propagate with an arbitrarily chosen group velocity and rotation. 

The study of ST helicon wavepackets can be treated within the framework of cylindrically symmetric waveguides, i.e., media with a radially dependent refractive index $n$. One may consider the bulk case as the limiting case of an infinitely large waveguide. In the following, we limit our study to a scalar approach by considering the weak guidance approximation. The partial differential equation governing the linear propagation of an electric field $E$ is then given in cylindrical coordinates by \cite{bejot2021spatiotemporal}:

\begin{equation}
    \left[ \partial^2_z+\partial^2_r+\frac{1}{r}\partial_r+\frac{1}{r^2}\partial^2_\theta+\frac{n^2(r,\omega)\omega^2}{c^2} \right] \Tilde{E}(r,\theta,z,\omega)=0,
    \label{eq:linearPDE}
\end{equation}

\noindent where $n(r,\omega)$ is the radial-dependent refractive index at $\omega$. $\tilde{E}$ is the Fourier transform of $E$ with respect to the time coordinate. Finding the ST wavepackets that are invariant under uniform rectilinear motion and rotation around the $z$ axis (but also invariant under phase transformation) amounts to finding the fields that belong to the kernel of the space-time screw axis symmetry differential operator $\Pi=\partial_z+K_1 \partial_t+K_l L_z-iK_0$, where $K_0$, $K_1$ and $K_l$ are arbitrarily chosen constants and $L_z=x\partial_y-y\partial_x=\partial_\theta$ is the $z$-component of the angular momentum operator. This can be done by expressing the field $\tilde{E}$ in the modal basis. In the case of a circularly symmetric medium (and in the weak guidance approximation), the modes correspond to fields that carry orbital angular momenta. The modal decomposition then reads as
\begin{equation}
    \tilde{E}(r,\theta,\omega,z)=\sum_{\ell,p}\bar{E}(\ell,p,\omega)F_{\ell,p}(r,\omega)e^{i\ell\theta}e^{iK_z(\ell,p,\omega)z},
\end{equation}
where $\ell$ and $p$ refer here to azimuthal and radial indices, where $\ell=(0,\pm1,\pm2,\pm3,...)$ is the topological charge, related to the phase front of OAM modes, $K_z(\ell,p,\omega)$ and $F_{\ell,p}(r,\omega)$ are the propagation constant and radial profile of the OAM mode $(\ell,p)$ at the frequency $\omega$, and $\bar{E}(\ell,p,\omega)$ is the decomposition coefficient of the field in the modal basis. 
In this basis, the linear evolution of the field along the propagation axis $z$ reads as	
\begin{equation}
    \partial_z\bar{E}(\ell,p,\omega)=iK_z(\ell,p,\omega)\bar{E}(\ell,p,\omega).
    \label{eq:linearevol}
\end{equation}
Looking for fields that belong to the kernel of $\Pi$ immediately implies that the decomposition of ST wavepackets only embeds a family of eigenvectors (i.e., modes) whose eigenvalues (i.e., propagation constants) obey the equation
\begin{equation}
    K_z(\ell,p,\omega_{\ell p})=K_0+K_1\omega_{\ell p}+K_\ell \ell,
    \label{eq:family}
\end{equation}
Accordingly, any electric field built from a given family is a diffraction- and dispersion-free ST wavepacket propagating at the group velocity $1/K_1$ whose intensity continuously rotates around the propagation axis with a spatial period $2\pi/K_\ell$ \cite{bejot2021spatiotemporal}.
Note that the propagation constant in a waveguide has a nonlinear dependence on the topological charge $\ell$. By only considering the second-order dispersion term around a given, arbitrarily chosen frequency $\omega_0$, the propagation constant of a waveguide can be well approximated as
\begin{equation}
    K_z^2\simeq\left(k_0+k_1\Omega+\frac{k_2}{2}\Omega^2\right)^2-\Gamma(\ell,p,\Omega),
    \label{eq:Kz}
\end{equation}
with $\Omega=\omega-\omega_0$ and $\Gamma$ a three-dimensional second-order polynomial: 
\begin{equation*}
\begin{split}
\Gamma(\ell,p,\omega) & = \gamma_{0,0,0}+\gamma_{1,0,0}\lvert \ell\rvert+\gamma_{0,1,0}p+\gamma_{0,0,1}\Omega \\
 & + \gamma_{2,0,0}\lvert \ell\rvert^2+\gamma_{0,0,2}\Omega^2+\gamma_{0,2,0}p^2\\
 & +\gamma_{1,1,0}\lvert \ell\rvert p+\gamma_{1,0,1}\lvert \ell\rvert\Omega+\gamma_{0,1,1}p\Omega 
\end{split}
\end{equation*}
where the coefficients $\gamma_{i,j,k}$ depend on the geometry of the considered waveguide. In the scalar case, the propagation constant $K_z$ does not depend on the sign of $\ell$ for a circularly symmetric waveguide. Inserting Eq.(\ref{eq:Kz}) in Eq.(\ref{eq:family}), we find that ST helicon wavepackets are contained in surfaces whose algebraic expressions are
\begin{equation} \label{eq:quadric} 
    \begin{split}
    \Gamma(\ell,p,\omega) & -(k_1^2+k_0k_2-K_1^2)\Omega^2+K_{\ell}^2\ell^2+2K_1K_{\ell}\Omega \ell \\
    & -2(k_0k_1-K_0K_1))\Omega+2K_0K_{\ell}\ell-(k_0^2-K_0^2)=0
    \end{split}
\end{equation}

As $\Gamma(\ell,p,\omega)$ depends on $\lvert \ell\rvert$, it is straightforward to notice that the surface embedding spatio-temporal wavepackets will be composed of two different quadrics (one for $\ell>0$ and another for $\ell<0$) with the same rank and nature. The complete classification of the nature of the quadric surfaces in the frequency-resolved modal space can be performed by studying the matrix associated to relation (\ref{eq:quadric}) and the relative signs of corresponding eigenvalues (see Ref. \cite{bejot2022quadrics}). Note that, for simplicity, we refer to and represent surfaces from relation (\ref{eq:quadric}), although only discrete values can be found for the parameters $(\ell,p,\omega)$. For each configuration, the total surface is composed of two different quadrics (one for $\ell>0$ and another for $\ell<0$), such as hyperboloids of one or two sheets. Furthermore, limited cases corresponding to conic sections (e.g., discrete X-waves) can be found when $\ell=0$ or $p=1$. Indeed, such surfaces encompass both helicon (monochromatic) beams \cite{vetter2014generalized} and discretized conical wavepackets \cite{zamboni2003superluminal,kibler2021discretized}, which are located at the intersection of the surface and planes of constant $\omega$ and constant topological charge $\ell$, respectively. Our theory then naturally encompasses previous concepts, which now appear as particular instances of spatio-temporal helicon wavepackets (e.g., ST light sheets \cite{kondakci2017diffraction}).

\begin{figure}
\begin{center}
\includegraphics[width=1\columnwidth]{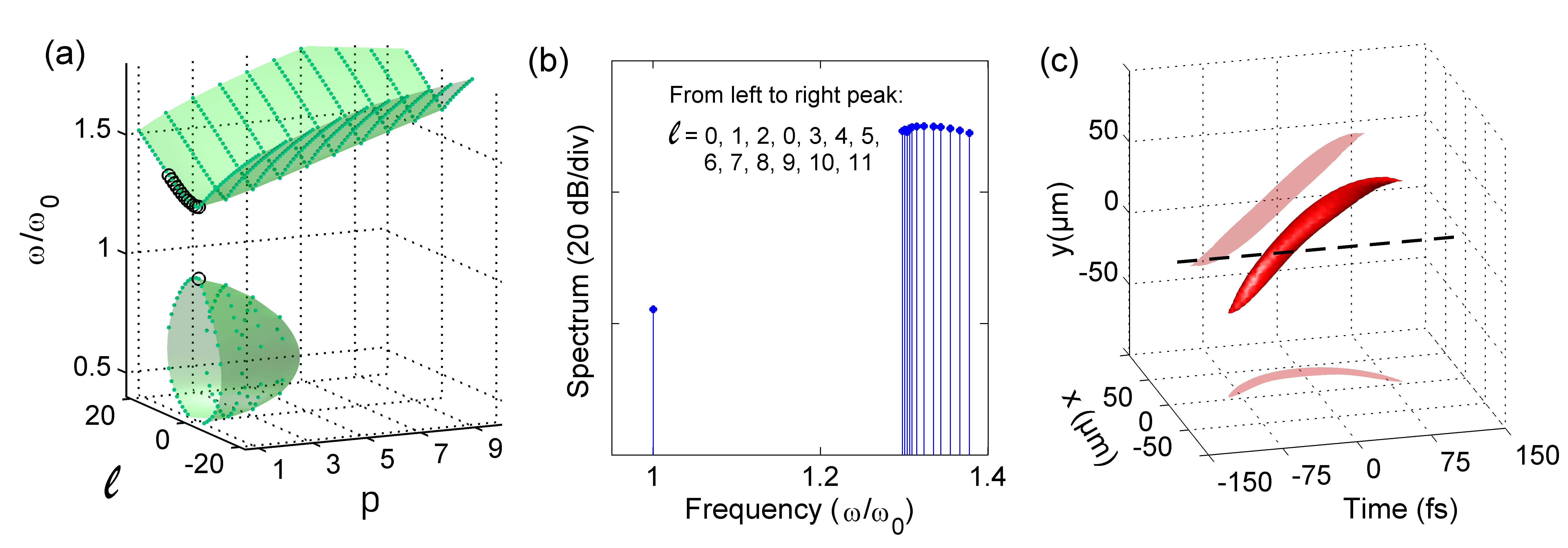}
\caption{Example of helicon wavepacket constructed from phase-matching of spatiotemporal components in a commercial MMF with core radius $R=$ 52.5 $\mu$m and numerical aperture $NA=$ 0.22. (a) Quadric surfaces obtained by using parameters of simulations in Sec. \ref{self-focus}. Green dots indicate discretized conic sections for each p coordinate. Black circles correspond to the selected modes used for the linear construction of helicon wavepacket. (b) Corresponding total spectrum of the selected modes. (c) Calculated iso-surface at half-maximum of the corresponding spatiotemporal intensity pattern of the constructed helicon wavepacket. The dashed black line indicates the origin $(x=0,y=0)$. Projections on planes (shadow plots) are also provided for a clear observation of rotating field pattern.}
\label{fig:linearhelicon}
\end{center}
\end{figure}

Next, we present an example that illustrates the general approach to synthesizing rotating ST wavepackets through the quadric phase-matching of their spatiotemporal optical components (Eq. \ref{eq:quadric}). Specifically, we construct an ST wave structure using the modes of a commercial MMF (used in the next sections): core radius $R=$ 52.5 $\mu$m (made of pure silica) and numerical aperture $NA=$ 0.22. For each $(\ell,p)$ couple, the roots of the quadric must be found, corresponding to the frequencies $\Omega_{\ell p}$ that represent the family of phase-matched modes $(\ell,p,\Omega_{\ell p})$ for given values of $K_0$, $K_1$, and $K_\ell$ . Figure \ref{fig:linearhelicon}(a) shows the overall three-dimensional pattern of phase-matching obtained in the normal dispersion regime, when considering a central wavelength of $\lambda_0=2\pi c/\omega_0=$ 1035\,nm, and arbitrary propagation constants, taken from the simulations which are described in the following section \ref{self-focus}. The linear construction of a helicon wavepacket then results from the superposition of the calculated frequencies $\Omega_{\ell p}$. 

Figure \ref{fig:linearhelicon}(b) presents an example of a particular case with selected modes that have specific radial and angular indices ($0\leq \ell\leq12$, and $p=1$), with spectral amplitudes governed by a gaussian shape and equal phases (see also the circles in Fig. \ref{fig:linearhelicon}a). The subplot (c) displays the spatio-temporal intensity iso-surface at half-maximum for a particular propagation distance. The wavepacket resembles a corkscrew in space-time coordinates with extreme localization, as all frequencies are in phase. It evolves with the spiraling trajectory of the intensity pattern, while maintaining its initial distribution during propagation. This helicon wavepacket does not disperse; it simply rotates around the $z$- and $t-$axes during propagation, owing to its invariant nature. The direction of rotation is driven by the sign of $K_{\ell}$. Careful selection of multiple higher-order modes at phase-matched discrete frequencies can easily produce complex engineered 3D-patterns.

\subsection{Self-focusing, shock dynamics and dispersive wave generation: MM-UPPE modeling}
\label{self-focus}

Multimode fibers are known to be versatile platforms for shaping supercontinuum generation through spatio-temporal oscillations and non-linear effects \cite{wright2015controllable,wright2015ultrabroadband}. In this section, we review the numerical modeling suitable for investigating the highly nonlinear supercontinuum generation regime in multimode fibers. More specifically, we focus on spatiotemporal helicon wavepackets, nonlinearly induced by appropriate and realistic initial pumping conditions \cite{bejot2021spatiotemporal}. Here, the numerical approach to highly non-linear pulse propagation is based on the MM-UPPE, recently derived in Ref. \cite{bejot2019multimodal}. Previous numerical models, such as the multimode generalized nonlinear Schrödinger equation and the Gross–Pitaevskii equation, were found to only provide qualitative support for the experimental results when a significant number of modes are involved (i.e., higher than 10) and ultrabroadband frequency conversion processes occur (i.e., larger than one octave) \cite{tarnowski2021numerical}. The first model considers a modal decomposition of the electric field, followed by intermodal nonlinear couplings. The second model is based on a direct representation of the electric field in the space–time domain; the refractive-index distribution acts as a potential term. In both approaches, strong approximations remain in the propagation equation. In contrast, the MM-UPPE follows a structure originating from the unidirectional pulse propagation equation, which is widely used to investigate extreme nonlinear optics in bulk transparent media and gases \cite{kolesik2004nonlinear}. Generally, approximate models tend to overestimate the overall nonlinear response compared to the MM-UPPE approach. This results in a broader spectrum and higher peak power in the space-time domain, as well as stronger power spectral densities in higher-order modes. It clearly demonstrates the significant influence of nonlinearity dispersion on pulse splitting and supercontinuum dynamics in the ultrashort pulse regime.

The MM-UPPE describes the evolution of the complex electric field $\xi$ in the scalar approximation as shown just below (expressed so that $\lvert \xi\rvert ^2=I(r,\theta,t)$, $I$ being the pulse intensity). The equation is written in a local frame propagating at an arbitrarily chosen velocity $v_{g_0}$, here equal to the group velocity of the fundamental mode $(\ell=0,p=1)$ calculated at $\omega_0$. 
\begin{equation} \label{eq:MMUPPE} 
       \partial_z \bar{\xi}=i\left(K_z-\frac{\omega}{v_{g_0}}\right)\bar{\xi}+i\frac{n_{\textrm{eff}_0} n_2 \omega^2}{c^2 K_z}\left[(1-f_R)\lvert \overline{\xi\rvert ^2 \xi} +f_R \overline{\left(\int h_R(\tau) \lvert \xi(t-\tau)\rvert^2 d\tau\right)\xi}\right],  
\end{equation}
where $\bar{A}$ denotes the frequency-dependent modal decomposition of an arbitrary space-time function $A(r,\theta,t)$, $n_{\textrm{eff}_0}$ is the effective refractive index of the fundamental mode at $\omega$, $n_2$ is the nonlinear refractive index of the medium (here for silica glass, we used $n_2=$ 3.2 $10^{-20}$ m$^2$/W). The function $h_R$ is the Raman response with fraction $f_R=$ 0.18 for fused silica glass \cite{agrawalnonlinear}. For simplicity, we have neglected here the nonlinear term responsible for third-harmonic generation, and the presence of usual fiber losses (less than 10 dB/km), which are found here to be negligible over the considered propagation distances. We recall that the present simulations are full 3D+1 simulations, and can be solved using a split-step algorithm based on the construction of a modal transform capable of rapidly switching from the space-time distribution of the field to its modal distribution \cite{bejot2019multimodal}. 

\begin{figure}
\begin{center}
\includegraphics[width=0.9\columnwidth]{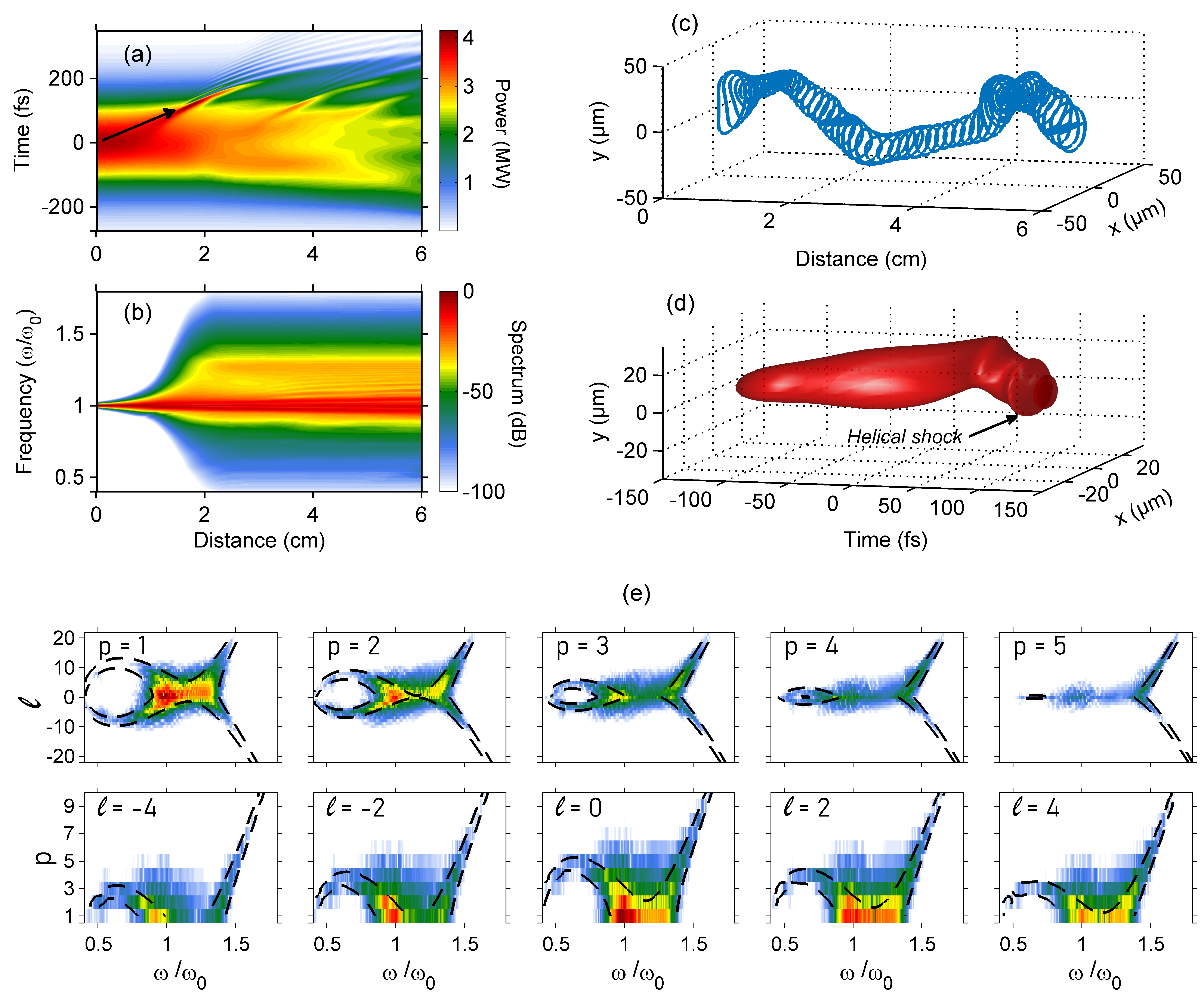}
\caption{Nonlinear propagation of 250 fs pulses (1.05 $\mu$J of energy at 1030 nm) in a 6-cm-long segment of step-index multimode fiber. (a,b) Evolution of the instantaneous power and normalized full power spectrum with propagation distance, respectively. The black arrow in (a) indicates both the location and the group velocity of the optical shock. (c) Evolution of the isocontour at the half-maximum of the fluence with propagation distance. (d) Isosurface at $1/e^2$ of the 3D intensity profile after 1.8 cm of propagation (optical shock distance). (e) Mode-resolved output spectra for the first radial (top panels) and angular (bottom panels) indices. The (dashed) black contours correspond to the phase-matching of theoretical radiations forming the spatiotemporal helicon wavepacket, propagating with a group delay of $\delta k_1$ and rotating around the propagation axis with an angular delay of $\delta k_\ell$ compared to the initial pump pulse centered at $\omega_0$. }
\label{fig:MMUPPE}
\end{center}
\end{figure}

As an example, we present the nonlinear propagation of 250-fs Gaussian pulse at $\lambda_0=$ 1030 nm, with 1.05-$\mu$J energy, in 6 cm of standard step-index multimode fiber (core radius $R=$ 52.5 $\mu$m and numerical aperture $NA=$ 0.22). Considering both radial and angular modes, our calculations include more than 1500 modes. The initial peak power is around 1.15 times the critical power of silica glass. Initially, the energy is equally distributed among the first two OAM modes of the fiber ($p= 0$, $\ell= 0$ and $+1$). The difference in the propagation constant between the OAM modes at $\omega_0$ is $K_z (1,1,0)-K_z (0,1,0)\simeq$ -178 rad m$^{-1}$. Accordingly, the laser pulse initially rotates around the propagation axis with a period of 3.5 cm, as shown in Fig. \ref{fig:MMUPPE}(c) by the contour-plot of fluence at half-maximum. Figures \ref{fig:MMUPPE}(a) and (b) show the detailed nonlinear propagation of both the temporal power and the power spectrum. We observe that the spectral dynamics reach a stationary state after 2 cm, just past the first pulse splitting dynamics observed in the time domain [see Fig. \ref{fig:MMUPPE}(a)]. Initial self-focusing dynamics, arrested by self-steepening, lead to the spontaneous formation of an intense optical shock (at the trailing edge) in the time domain, as well as strong spectral broadening. The shock front moves in the same direction in the retarded time frame, with a group velocity that is nearly uniform (but different from $v_{g_0}$), as indicated by the black arrow in Fig. \ref{fig:MMUPPE}(a). The shock is characterized by a relative group delay of 7\,ps\,m$^{-1}$. The simultaneous strong broadening associated with this optical shock generates linear waves that are resonantly amplified in higher-order modes. A clear structure emerges in the mode-resolved spectrum at the distance at which this optical shock occurs, in the form of a discretized conical wave (also known as X-wave) over angular indices $\ell$ for low radial indices $p$, as depicted in Fig. \ref{fig:MMUPPE}(e). This particular shape corresponds to the formation of a spatiotemporal helicon wavepacket belonging to a specific mode family, as defined in the previous section. In other words, it corresponds to the linear superposition of phase-matched resonant radiations over the fiber OAM modes. More specifically, the emerging helicon wavepacket propagates and rotates according to the velocity properties of the shock front. 

The underlying approach to interpreting the generated OAM-carrying supercontinuum is similar to the effective three-wave mixing model previously developed to describe broadband spectral broadening and generation of conical waves in bulk media \cite{kolesik2005interpretation}. The nonlinear optical response of the medium is driven by the strongly localized optical wave packets which are formed during space-time pulse propagation. These wavepackets are scattered and seed linear waves at new frequencies, as dictated by a complex phase-matching relation. Helicon wavepackets are defined by two parameters, namely, their propagation velocity and their rotation period, which are related to their spiraling trajectory of superposed OAM fiber modes. Accordingly, their energy is concentrated in families of modes $(\ell,p,\omega_{\ell p})$ that here propagate and rotate at the same rates as the shock front (SF):
\begin{equation}
    \lvert K_z(\ell,p,\omega_{\ell p})-K_{0,SF}-K_{1,SF}\omega_{\ell p}-K_{\ell,SF}l \rvert \leq 2\pi/d_r
  \label{eq:phasematching}  
\end{equation}
where $K_{0,SF}$, $K_{1,SF}$, and $K_{\ell,SF}$ are the propagation constant, the inverse of group velocity, and the rotation rate of the shock front, respectively. Since the nonlinear process takes place over a finite length $d_r$, Eq. (\ref{eq:phasematching}) does not represent a strict phase-matching condition, and it can be satisfied within a certain tolerance related to $d_r$ \cite{kibler2021discretized,kolesik2005interpretation}. Together with the precise knowledge of the frequency-dependent propagation constants, Eq. (\ref{eq:phasematching}) then allows the shape of the mode-resolved output spectrum shape of the pulse impacted by helicon wave emission to be estimated. $K_{0,LW}$ is equal to the propagation constant of the input pump pulse $K_0(l=0,p=1,\omega_0)$. From Fig. \ref{fig:MMUPPE}(a) and the black arrow, the inverse of the group velocity is found to be $K_{1,LW}= 1/v_{g_0}+\delta k_1$, with $\delta k_1=$ 7 ps m$^{-1}$. The rotation rate $K_{\ell,LW}$ is also impacted by nonlinear propagation, in a similar way to the group velocity. This is corroborated by the 3D-intensity profile which is spiraling in time, as obtained at the distance of conical wave emission, and depicted in Fig. \ref{fig:MMUPPE}(d). The isosurface of the intensity clearly exhibits an optical shock spiraling, characterized by different positions and shapes in space, thus suggesting a variation of rotation feature along the pulse splitting. In Ref. \cite{bejot2021spatiotemporal}, a detailed analysis has shown that the intensity peaks do not rotate at the same velocity as the initial propagation constant difference between the OAM modes at $\omega_0$. Thus, the rotation rate of the localized wave can be determined as $K_{\ell,LW}= K_z(1,1,0)-K_z(0,1,0)+\delta k_\ell$, where in the present case $\delta k_\ell\simeq$ 11 rad m$^{-1}$. Figure \ref{fig:MMUPPE}(e) shows the angularly-resolved spectra for the five first radial and angular indices obtained from numerical simulation at the output of the fiber. The dashed black iso-contour superimposed to the mode-resolved output spectrum delimits the inner angular-spectral region that satisfies Eq. \ref{eq:phasematching} with the previously fixed parameters, when $d_r=$ 4\,mm (i.e., seeding distance of the helical shock). All the mode-resolved output spectral shapes are well fitted by this contour, which confirms that a helicon wavepacket rotating at $K_{\ell,LW}$ and propagating at group velocity $1/K_{1,LW}$ is generated during nonlinear propagation. A clear conical wave of X-wave type is revealed over the lowest orders of OAM fiber modes, a typical signature of pumping in the normal dispersion of the medium or waveguide.

\subsection{Experimental demonstration: spontaneous emergence of discretized conical wave}
\label{X-wave}

\begin{figure}
\begin{center}
\includegraphics[width=0.7\columnwidth]{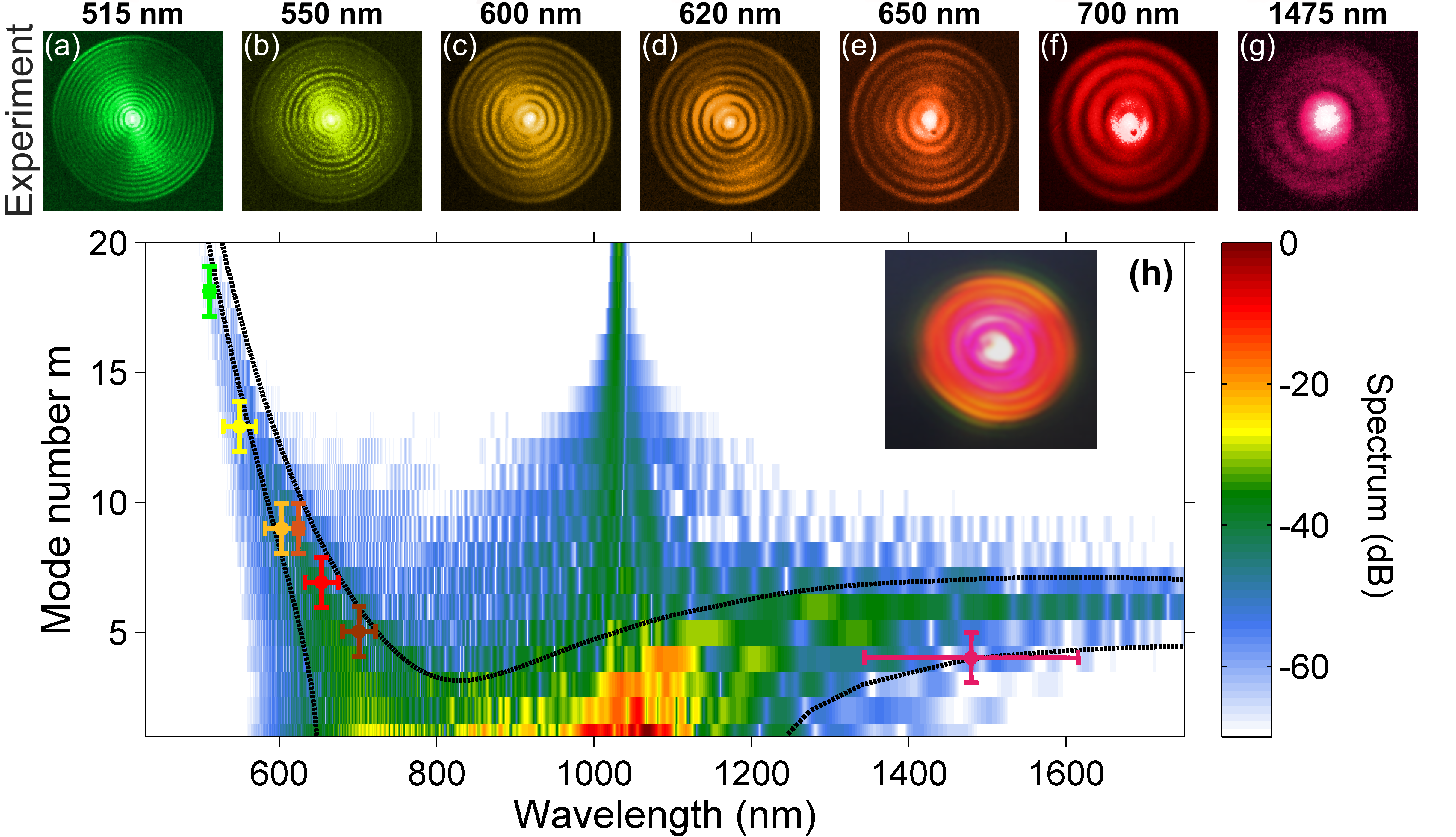}
\caption{Mode-resolved spectrum of conical emission obtained at the output of the MMF. (a-g) Near-field images of the MMF higher-order modes involved in both X-like tails at specific wavelengths (experimental captures obtained with CMOS sensor). (h) False-color map of the numerical mode-resolved spectrum obtained with MM-UPPE. Black dotted lines represent the theoretical phase-matching region of the discretized conical wave emitted during the optical shock. Colored crosses indicate spectral positions of the higher-order modes captured in subplots (a)-(g) according to the bandpass spectral filters used. Inset: photo of the far-field pattern recorded at the output of the MMF.}
\label{fig:Xwavedemo}
\end{center}
\end{figure}

Discretized conical emission in a MMF has been demonstrated using a high-power femtosecond laser at 1035 nm with $\mu$J-level energy, which was carefully coupled into the fundamental mode of a commercial step-index silica fiber (5.8\,cm-long segment) with a 105-$\mu$m pure silica core \cite{stefanska2023experimental}. The spontaneous emergence of a discretized conical wave of X-type has been analyzed from theory to experiments. This reveals that this spatiotemporal phenomenon corresponds to broadband intermodal dispersive wave emission from an unsteady localized wave structure formed during nonlinear propagation. A distinctive spectral feature indicating the formation of a conical wave was observed as a broad and strong spectral broadening toward visible wavelengths (i.e., blue-shifted spectral shoulder associated to the formation of an optical shock). Additional bandpass filters were also included in the detection system, to characterize the modes associated with newly generated frequencies. In Ref. \cite{stefanska2023experimental}, the discretized conical wave was also characterized in terms of angularly resolved far-field spectra. Similar features were unveiled in the process of typical (X-wave) conical emission obtained in a sapphire plate. Beyond the continuous spectral broadening on-axis, the energy spreading follows an increasing cone angle of emission with increasing detuning from the carrier wavelength. Figure \ref{fig:Xwavedemo} shows the numerical results of the distribution of the full optical spectrum (power in log scale) over the different fiber modes after propagation. Experimentally, the modal composition of both visible and infrared X-like tails was characterized by means of near-field imaging of the MMF
output facet, combined with bandpass spectral filtering [see subplots (a-g)]. Increasingly detuned frequencies from the pump are contained in higher-order modes of the MMF (e.g., emission in LP$_{0,18}$ at 515\,nm) and corroborate the mode-resolved spectrum obtained from MM-UPPE simulation. 

This normal-dispersion scenario of conical emission occurs near the nonlinear spatiotemporal focus, owing to the self-steepening effect. The spectral broadening associated with shock formation at the
trailing edge of the pulse clearly seeds a large number of modes over a finite distance, but only phase-matched dispersive radiations over specific higher-order modes remain with significant intensities according to the velocity of the shock front, as described by Eq. \ref{eq:phasematching} (here only LP$_{0,m}$ modes were considered). Consequently, the parameters $\delta k_1=$ 12.6\,ps.m$^{-1}$ and $d_r=$ 1.4\,mm determined from simulation provide an excellent interpretation of phase-matching as shown in Fig. \ref{fig:Xwavedemo}h (see black dotted lines).

\subsection{Optimizing ultrabroadband supercontinuum generation}
\label{SCgen}

Nonlinear pulse propagation in multimode waveguides is strongly dependent on spatiotemporal couplings \cite{wright2022physics}. To maximize spectral broadening and control the spatial content of the generated supercontinuum (SC), initial coupling into the waveguide's fundamental mode might be required for simplicity, allowing an adiabatic change of the mode and minimizing energy loss. This prevents detrimental effects such as numerous intermodal couplings with low modal overlap for nonlinear effects, and the filtering of higher-order non-guided modes in nonuniform waveguides (e.g., tapered fibers). However, even in the ideal scenario of fundamental mode excitation combined with femtosecond pulses and short propagation, spontaneous intermodal phase matching can be observed, as shown in the previous section, when peak powers that imply self-focusing are used. It is then crucial to find the optimal coupling condition in such multimode glass rods and favor frequency conversion processes in the fundamental guided mode.

Numerical simulations in longitudinally uniform multimode fibers (see the previous section) have shown that, even around the self-focusing threshold, most of the total energy can remain in the initially excited mode (typically around 80-90 $\%$). The generated linear space-time wave packet is only part of the complex space-time dynamics, in which shock formation and pulse splitting predominate. In this context, the concept of a tapered multimode fiber has been proposed and demonstrated to enhance spectral broadening in the high-power regime \cite{serrano2025towards}. Particular emphasis has been placed on the mid-infrared region, where a new standard of fiber-based SC has been established. Different types of glass fiber have also been successfully used across various spectral windows. Simple postprocessing of glass rods (single-index fibers) offers new possibilities in terms of coupling efficiency, spectral coverage, and output power for supercontinuum design. The dimensions of the taper profile are highly important; the waist dimension, for example, has a strong impact on nonlinear spectral broadening, while the transition characteristics influence linear losses. The nonlinear coefficient increases significantly for the smallest rod diameters in the waist section, while the fundamental mode exhibits large effective areas at the input and output ends, facilitating high coupling efficiencies. This approach typically makes use of simple, easy-to-align, long-focus coupling lenses. The SC spatial profile exhibits a cleaning effect at the output of the tapered MMF, which is induced by higher-order mode filtering in the tapered section \cite{serrano2023multi}. Furthermore, simplified numerical simulations have confirmed the significant role of the fundamental mode in spectral dynamics\cite{serrano2025towards}. 

The most notable experimental demonstration is the unlocking of ultrabroadband supercontinuum generation in the mid-infrared and beyond 10\,$\mu$m wavelength, combined with high average power operation using commercial mid-IR laser sources with MHz-rate frequencies. SC generation in short segments of tapered glass rods (i.e., tapered MMFs), each a few centimeters long and made of single-glass materials, has been demonstrated \cite{serrano2025towards}. Only a moderate ratio of tapering (typically a factor of 4–5) is generally required. Using selenium-rich glass rods, it was possible to cover the 2–15\,$\mu$m wavelength range with a total average power of more than 100\,mW, thus providing crucial average power for spectroscopy measurements at wavelengths where it is currently difficult to do so \cite{bailleul2025integrated}.

\begin{figure}
\begin{center}
\includegraphics[width=0.75\columnwidth]{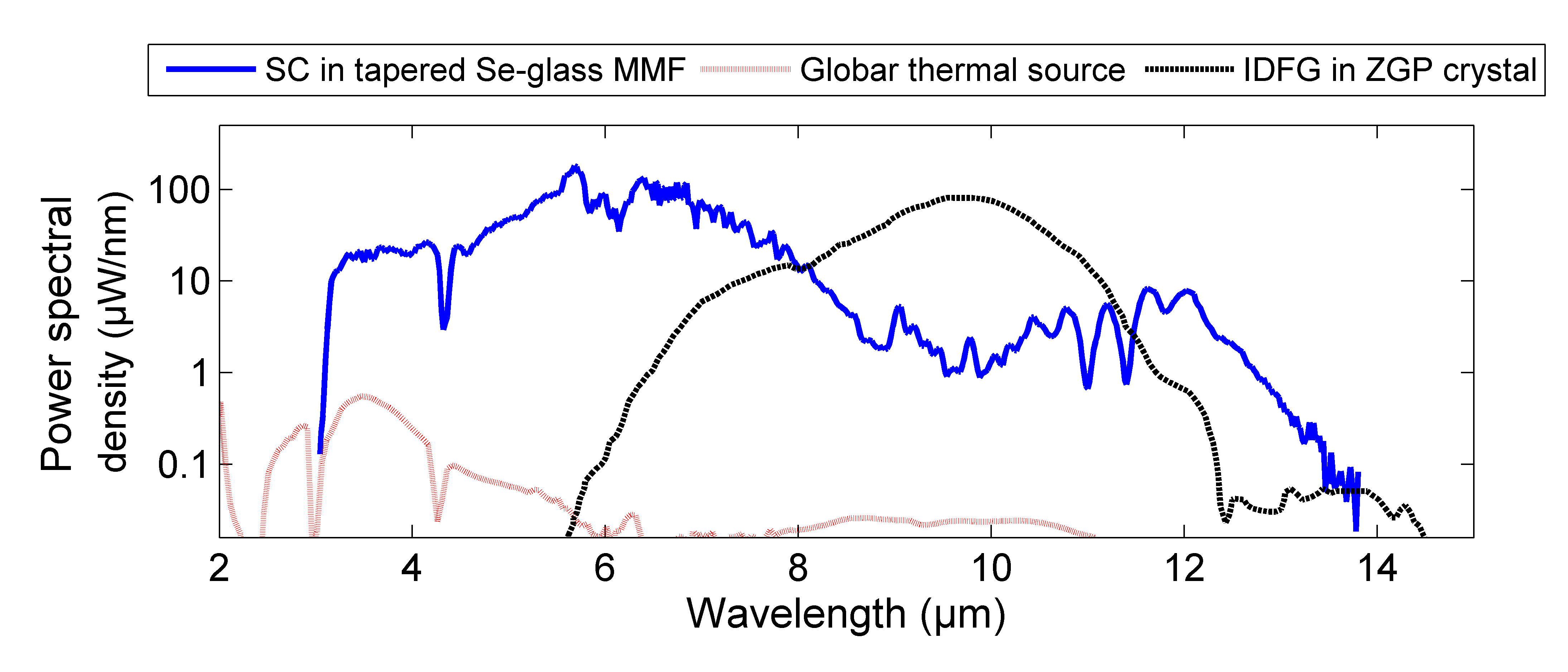}
\caption{Mid-IR SC source based on a tapered Se-glass MMF pumped by a commercial mid-IR laser chain compared with an example of super-octave mid-IR electromagnetic transients via intra-pulse difference frequency generation (IDFG) obtained in nonlinear quadratic crystals pumped by high-power mid-IR few-cycle pulses \cite{vasilyev2019super}. IR blackbody radiation collected from a commercial SiC globar is also presented for comparison of the power spectral density.
}
\label{fig:SCgen}
\end{center}
\end{figure}

These devices offer a promising solution by surpassing the limitations associated with (nearly single-mode) step-index or microstructured fibers, such as a low damage threshold and power coupling restrictions due to small mode areas. In the context of recent developments in broadband ultrafast laser spectroscopy, these fiber-based SC sources with a high power spectral density can already compete with the latest advances in broadband mid-IR spectra obtained from nonlinear quadratic crystals via intra-pulse difference frequency generation, which is pumped by high-power mid-IR few-cycle pulses (see Figure \ref{fig:SCgen}).

\vspace{1cm}
\section{Extreme nonlinear optics in hollow-core photonic crystal fibers}
\label{sec:limoges}
\subsection{Guidance principles of HCPCF}
\begin{figure}[H]
\begin{center}
\includegraphics[width=0.88\columnwidth]{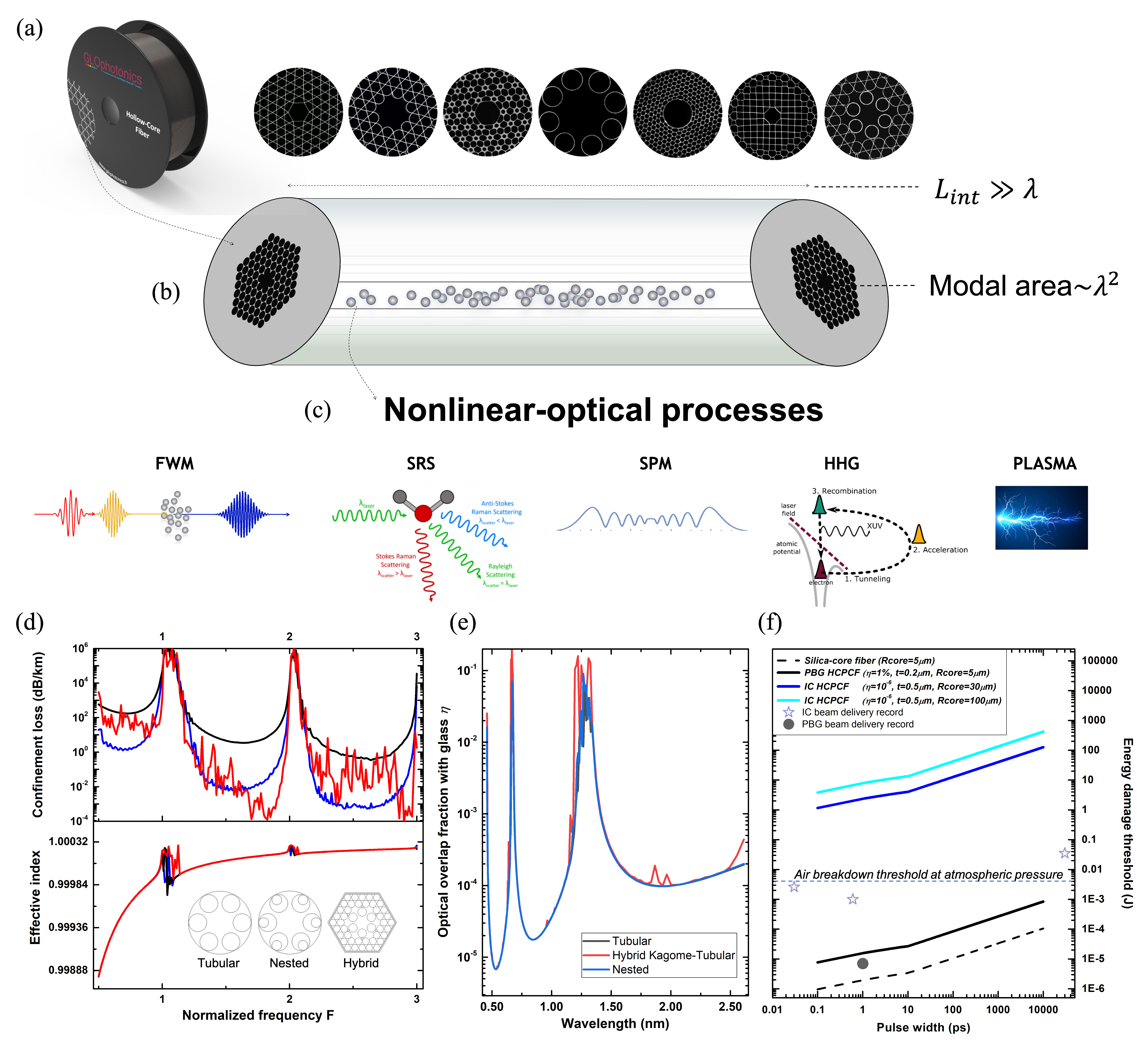}
\caption{(a) Examples of commercially available HCPCF. (b) Artistic illustration of a hollow-core fiber functionalized by the insertion of a gaseous medium and (c) the variety of nonlinear-optical processes involved. Computations made on three representative state-of-the-art HCPCF designs based on tubular, hybrid and nested cladding (core diameter 34\,$\mu$m, gap between holes 4.8\,$\mu$m and silica thickness 641\,nm) : evolution of the confinement loss and n$_{\textrm{eff}}$ (d) and the optical overlap with silica glass (e). (f) HCPCF-guided laser energy damage threshold for different types of HCPCF. The star and square symbols are the experimental demonstrated
energy with HCPCF-guided lasers based on inhibited coupling (IC) \cite{Gerome50,Gerome30,Gerome49} and bandgap guidance (PBG) \cite{Gerome48}, respectively. The air ionization threshold at 1 bar is shown for indicative purposes. 
}
\label{fig:HCPCF_intro}
\end{center}
\end{figure}

The advent of HCPCFs \cite{cregan1999single} has provided a platform that has profoundly impacted distinct disciplines: telecommunications, nonlinear and quantum optics. HCPCF is a class of microstructured fibers whose hollow-core guidance and design versatility underpin their extraordinary nonlinear-optical (NLO) performance. As shown in \cref{fig:HCPCF_intro}(a), a variety of commercially available geometries—ranging from single-ring tubular and hybrid Kagome structures to nested tubular designs—illustrate the wide freedom in cladding architecture, guidance mechanisms, and core diameters. This structural flexibility allows dispersion, confinement loss, and modal area to be engineered across several orders of magnitude, making HCPCFs a truly universal photonic platform bridging the gap between free-space propagation and conventional solid-core fibers. The consequence of this design flexibility is vividly reflected in the diversity of nonlinear phenomena summarized in \cref{fig:HCPCF_intro} (b-c), where gas-filled HCPCFs have enabled an unprecedented range of effects: stimulated Raman and Brillouin scattering, self-phase modulation, soliton self-compression, dispersive-wave emission, supercontinuum generation from the UV to the mid-IR, plasma formation, high-harmonic and attosecond pulse generation, and even quantum-light processes such as photon-pair creation and frequency conversion. Together, these results define the field now termed \textit{gas photonics} \cite{Benabid1}, where light, gas, and glass act as a single synergistic system to access regimes of nonlinearity, coherence, and field strength that are unreachable in bulk media or solid fibers. The exceptional guiding characteristics of HCPCFs provide the physical basis for this wealth of phenomena. 

\Cref{fig:HCPCF_intro}d illustrates their salient features: confinement losses can reach values below $10^{-4}$ dB/km, while chromatic dispersion remains extremely low and tunable via structure and gas pressure. Equally important, the fraction of optical power overlapping with the surrounding silica—shown in \cref{fig:HCPCF_intro}(e) can be as small as $\eta$ $\sim$ $10^{-6}$, effectively isolating the guided light from the glass and granting unprecedented energy-handling capability. Consequently, as shown in \cref{fig:HCPCF_intro}(f), the laser-damage threshold of HCPCFs exceeds that of solid silica by several orders of magnitude; even femtosecond or picosecond pulses with energies approaching the joule level can, in principle, propagate without inducing structural degradation. To put this into perspective, a 1–10\,J energy handling means that a 10 fs pulse confined in a 100\,$\mu$m core would corresponds to peak intensities of $10^{17}$–$10^{18}$\,W/cm$^{2}$, approaching the onset of the relativistic-optics regime (typically above $10^{18}$\,W/cm$^{2}$ for near-infrared wavelengths)\cite{RevModPhys.91.030501}, thus positioning gas-filled HCPCFs as an arguably promising waveguide for relativistic optics and compact platforms for guided extreme-field phenomena.

These capabilities explain why gas-filled HCPCFs behave as fiber-integrated gas cells capable of supporting extreme light–matter interaction over macroscopic distances. The combination of small modal area, long interaction length, and high damage threshold has fundamentally reshaped nonlinear fiber optics, giving rise to what is now recognized as extreme nonlinear optics (Extreme NLO). The chronological progression of discoveries that unveiled and exploited these features is summarized in Table 1, which lists the seminal experimental milestones in nonlinear and quantum optics achieved with HCPCFs. Each of these results not only demonstrated a unique nonlinear phenomenon but also revealed a fundamental property of HCPCFs—be it dispersion control, energy confinement, or interaction engineering—that continues to define their scientific and technological potential.
In the following sections, we review the major advances in nonlinear optics achieved with HCPCFs since their advent, highlighting both the fundamental insights and the transformative applications that have emerged from this unique platform.

\newcolumntype{Y}{>{\RaggedRight\arraybackslash}X}
\newcolumntype{S}{>{\RaggedRight\arraybackslash\hsize=0.3\hsize}X} 

\begin{table}[H]
\centering
\small
\begin{tabularx}{\textwidth}{c c Y S}
\toprule
\textbf{Year} & \textbf{Authors / Reference} & \textbf{Effect / Breakthrough} & \textbf{Gas medium} \\
\midrule
2002 & Benabid et al.\cite{Benabid3} & First demonstration of stimulated Raman scattering in H$_2$-filled HCPCF & H$_2$ \\
2003 & Ouzounov et al.\cite{ouzounov2003generation} & First soliton generation in a gas-filled HCPCF & Xe \\
2004 & Benabid et al.\cite{Benabid4} & Pure rotational Raman conversion with $>$90\% efficiency & H$_2$ \\
2007 & Couny et al.\cite{Benabid5} & Multi-octave Raman frequency comb generation & H$_2$ \\
2009 & Heckl et al.\cite{Benabid12,Gerome45} & First high-harmonic generation in Xe-filled HCPCF & Xe \\
2010 & Wang et al.\cite{Benabid6} & Quantum-noise-seeded phase-locked Raman comb generation & H$_2$ \\
2011 & Hölzer et al.\cite{Benabid9} & Experimental observation of plasma-driven soliton blue shift & Ar \\
2013 & Debord et al.\cite{Benabid19} & Microwave-driven plasma generation and confinement inside HCPCF & Ar-plasma \\
2015 & Balčiūnas et al.\cite{Benabid13} & Single-cycle pulse compression using gas-filled HCPCF & Xe \\
2015 & Finger et al.\cite{Benabid16,Benabid17} & Generation of twin-beam squeezed vacuum—first quantum light in HCPCF & Ar \\
2015 & Belli et al.\cite{Gerome22} & Vacuum ultraviolet supercontinuum generation in HCPCF & H$_2$ \\
2016 & Alharbi et al.\cite{Benabid15} & Molecular self-organization and SRS in the Dicke regime forming a nanolattice & H$_2$ \\
2018 & Debord et al.\cite{Benabid11} & Supercontinuum generation in air-filled HCPCF & Air \\
2019 & Cordier et al.\cite{cordier2019active,Benabid32} & Versatile photon-pair generation and high-SNR heralded single photons & Xe \\
2019 & Okaba et al.\cite{okaba2019superradiance} & Superradiance from lattice-confined atoms inside HCPCF & Cold Sr \\
2022 & Tyumenev et al.\cite{tyumenev2022tunable} & Raman-driven tunable and state-preserving single-photon frequency conversion & H$_2$ \\
\bottomrule
\end{tabularx}
\caption{Seminal experimental results in nonlinear and quantum optics using gas-filled HCPCFs.}
\label{Tab1}
\end{table}

\subsection{Stimulated Raman scattering (SRS)}
\begin{figure}[H]
\begin{center}
\includegraphics[width=1\columnwidth]{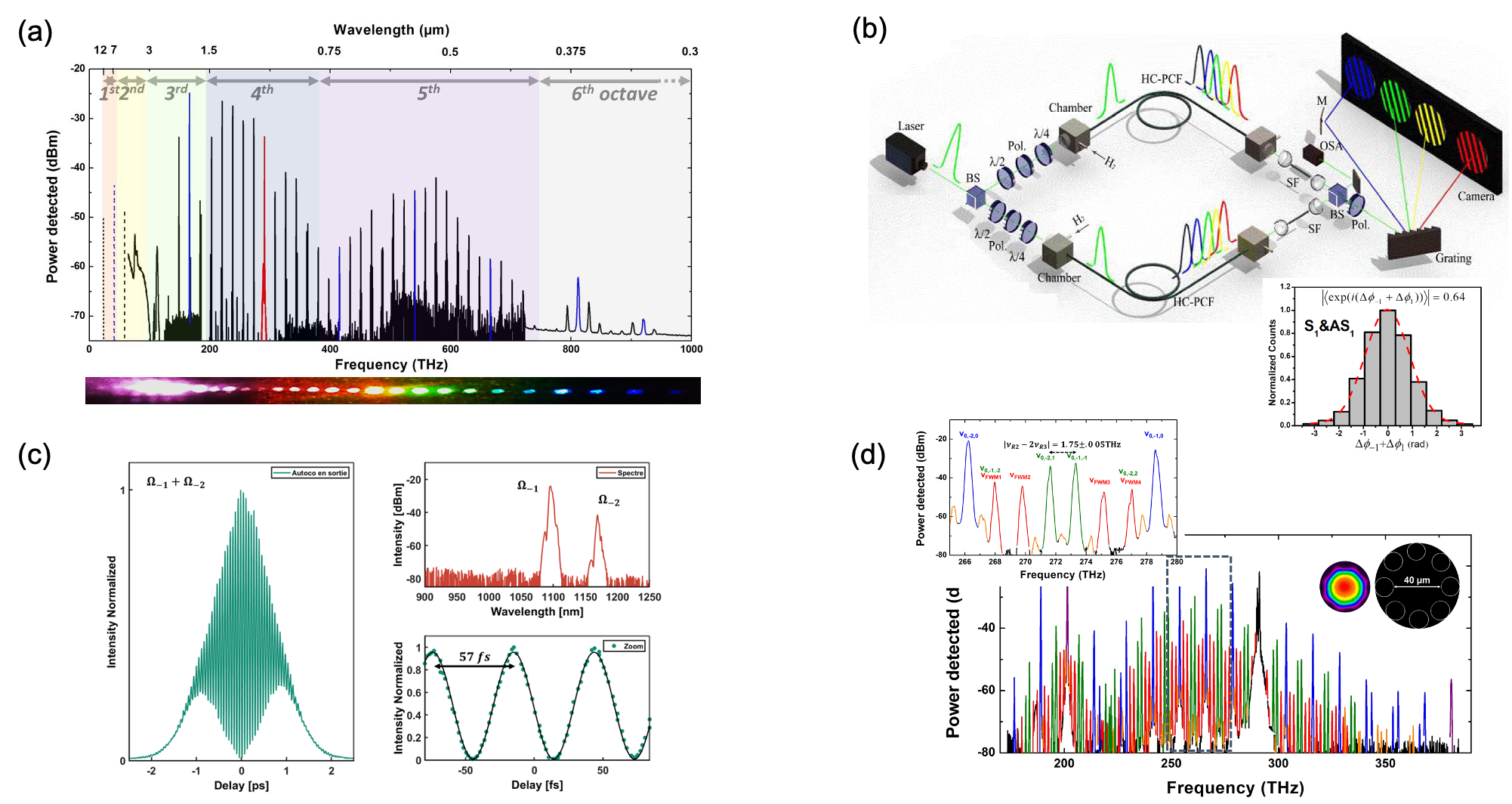}
\caption{(a) Raman frequency comb spectrum generated over 5 octaves in a 3 m-long hydrogen-filled HCPCF with a 27 ps laser pump linearly polarized (data from \cite{Gerome46}). (b) Interfering photons generated from independent quantum events (data from \cite{Benabid6}).(c) Temporal trace of the 1$^{\textrm{st}}$ Stokes / 2$^{\textrm{nd}}$ Stokes line at $\Omega_{-1}+\Omega_{-2}$. The inset shows a zoom-in and the optical spectrum. (d) Output spectrum after a 16 W pump propagation in a 3 m-long HCPCF filled with 40 bar of deuterium. Zoomed-in spectrum of the black-dashed rectangle showing the Stokes and anti-Stokes lines due to ortho-rovibrational Raman resonance by blue, para-rovibrational Raman resonance by green, and four-wave-mixing generated sidebands by red (data from \cite{Benabid7}).} 
\label{fig:HCPCF_SRS}
\end{center}
\end{figure}

The first breakthrough demonstration of SRS in HCPCF was reported in 2002, where hydrogen-filled fibers showed remarkably low threshold Raman conversion at sub-microjoule levels \cite{Benabid3}. This was followed by demonstrations of pure rotational Raman lasing in hydrogen by suppressing the vibrational Raman conversion thanks to the limited fiber transmission bandwidth \cite{Benabid4}, achieving conversion efficiencies exceeding 90\%. One of the most influential developments was the generation of multi-octave Raman combs in gas-filled HCPCFs. Couny et al. \cite{Benabid5} demonstrated cascaded Raman sidebands spanning more than 3 octaves in hydrogen-filled HCPCF by using only a few nanoseconds pump pulses. Later, they extended this to over five octaves by injecting a picosecond pulsed fiber laser \cite{Gerome46} (see \cref{fig:HCPCF_SRS}a). Further studies revealed that these combs are not random but phase-locked, with their coherence seeded by quantum noise in transient SRS \cite{Benabid6,Gerome47} (see \cref{fig:HCPCF_SRS}(b,c)). A particularly rich regime arises when Raman and Kerr nonlinearities couple, leading to the formation of Raman-Kerr combs \cite{Benabid7}. In this regime, the effective Kerr nonlinearity induced by Raman coherence enables quasi-equidistant frequency grids with much smaller spacing than individual Raman transitions. This has been achieved, for example, in deuterium-filled HCPCFs, where Raman–Kerr interactions produced dense combs with terahertz spacing (\cref{fig:HCPCF_SRS}d). Similarly, using a gas mixture of hydrogen, deuterium, and xenon, Hosseini et al. generated a frequency comb spanning from 280 nm to 1 $\mu$m, with a spacing of 2.2\,THz (135 lines) using a nanosecond laser at 532\,nm and pulses of 5\,$\mu$J \cite{Gerome1}. These results illustrate how HCPCFs serve as platforms not merely for more efficient SRS, but for qualitatively new spectral phenomena that bridge Raman and Kerr physics.

More recent results have also led to significant advances in the UV range, which has strong applicative interest in fields such as biology and chemistry. Mridha et al. \cite{Gerome2} demonstrated a broad Raman comb in the UV by pumping with a 532 nm laser combined with a 266 nm UV laser in a hollow-core fiber filled with hydrogen, achieving a conversion efficiency close to 60\%. These same authors also studied SRS of the backscattered wave in gases \cite{Gerome3}, which offers a promising method for compressing and amplifying laser pulses. They observed this phenomenon for the first time in a hydrogen-filled Kagome-type hollow-core fiber at high pressures near 100 bar. Conversion efficiencies exceeding 40\% were achieved for generating backscattered light from a 532 nm pump. The efficiency even reached 65\% when this process was amplified by the co-propagating Stokes wave, which was reinjected at the fiber output using a reflective mirror. At high pump powers, the backscattered signal proved stronger than its forward-propagating counterpart.
In the same vein, Tyumenev et al. \cite{Gerome4} obtained a UV comb extending down to wavelengths as short as 141 nm. Their approach involved using two pumps—one in the green and one in the UV—in a hydrogen-filled hollow-core fiber to extend the comb.
Finally, it is worth noting that research on frequency combs may also find new applications in quantum optics. Raman combs could provide a pathway for frequency conversion while preserving quantum correlations. In 2022, Tyumenev et al. \cite{tyumenev2022tunable} reported a pioneering result in this direction, demonstrating efficient preservation (up to 70\%) of the quantum state of photons through up-conversion and molecular modulation in hydrogen-filled hollow-core fibers. By combining a high-intensity coherent pump pulse with a low-intensity quantum source based on parametric down-conversion (generating signal/idler photon pairs), the authors showed that idler photons could be frequency-shifted by 125 THz via the Raman scattering process. This method offers the advantage of generating photons while preserving their quantum state across a broad and easily tunable spectral range, making it adaptable to other quantum sources. 

These results illustrate how HCPCFs serve as platforms not merely for more efficient SRS, but for qualitatively new spectral phenomena that bridge Raman and Kerr physics.

\subsection{Supercontinuum generation (SC)}
SCG in HCPCFs has become one of the most emblematic demonstrations of extreme nonlinear optics. In traditional solid-core fibers, SCG relies on Kerr-driven processes such as self-phase modulation and soliton fission. In contrast, gas-filled HCPCFs allow dispersion engineering via gas pressure and introduce ionization effects that enrich the dynamics \cite{Benabid8}. Moreover, the ultrabroadband guidance offered by the SCG mechanism makes HCPCFs uniquely suited for these nonlinear applications.

\begin{figure}[H]
\begin{center}
\includegraphics[width=1\columnwidth]{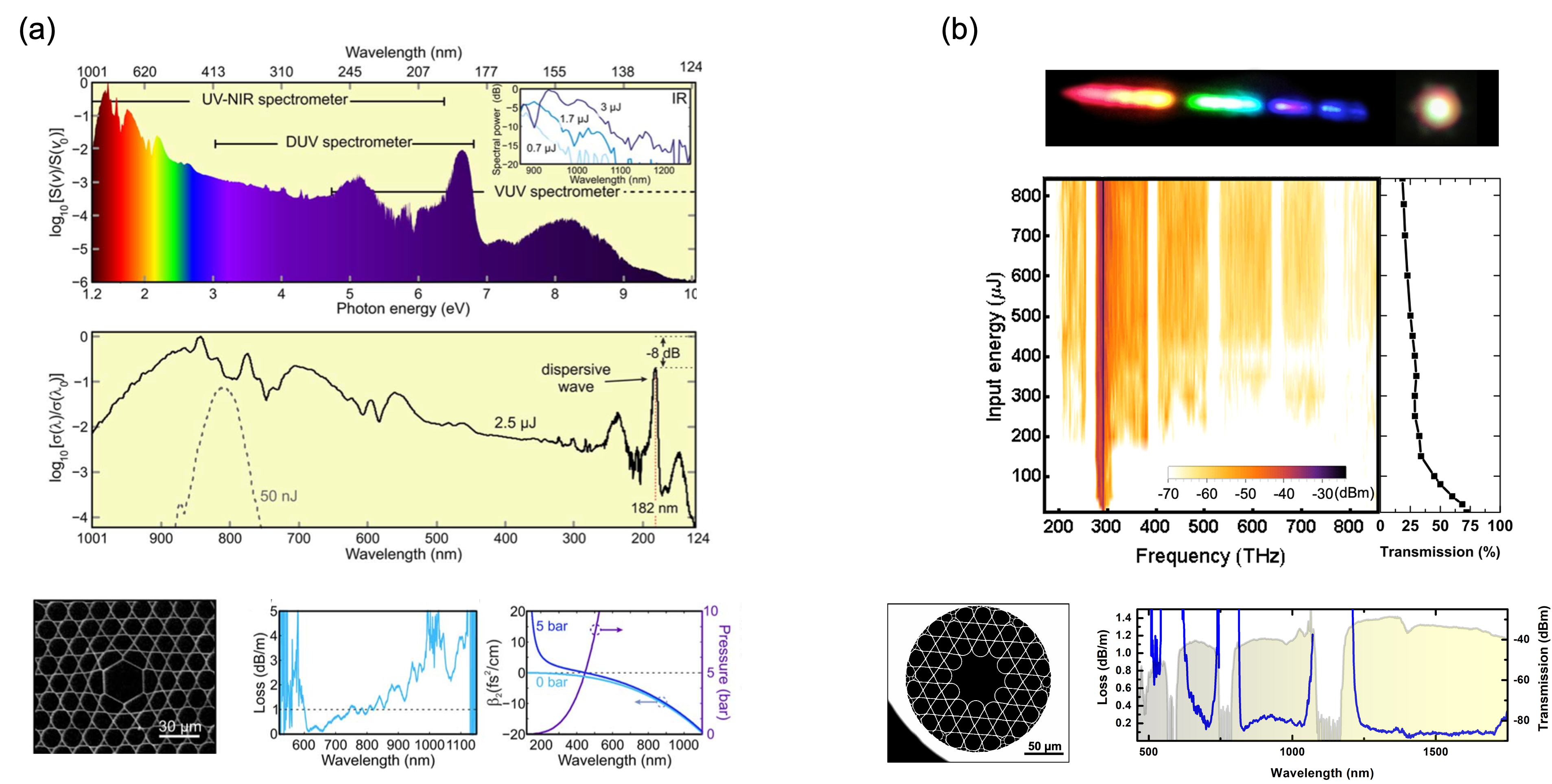}
\caption{(a) Results from \cite{Gerome22}  : Experimental SC spectrum generated by 30-fs pulses at 805 nm with an energy of 2.5 µJ, measured after propagation through 15 cm of a 5 bar hydrogen-filled HCPCF. (b) Experimental evolution of the output spectrum from a 3.8 m-long air-filled HCPCF for laser pulse energies up to 840 µJ and 600 fs duration at 1030 nm. (top) Dispersed beam of the generated SC at the fiber output (data from \cite{Benabid11}). The SEM images of the cross section of the fibers used and the corresponding loss curves for the two experiments are also indicated at the bottom. 
}
\label{fig:HCPCF_SC}
\end{center}
\end{figure}

For example, experiments have shown SC ultrabroad spectra spanning from vacuum UV (124 nm) to beyond 1200 nm in a hydrogen-filled fiber (see \cref{fig:HCPCF_SC}a) \cite{Gerome22} and then extented to the mid-infrared region ($>$5\,$\mu$m) when using mid-IR pump laser \cite{Gerome21}. The key mechanisms involve soliton self-compression in the anomalous dispersion regime, dispersive wave emission in the UV \cite{Gerome51}, and plasma-induced nonlinearities (detailed discussions will be presented in the section on photoionization-induced plasma) that lead to unique features such as the soliton blue-shift \cite{Benabid9,Benabid10}—the plasma analogue of the well-known Raman red-shift. These combinations were later optimized to generate a SC extending down to 113 nm through soliton-ionization dynamics in helium-filled fibers \cite{Gerome23}. More recently, the short-wavelength limit of SCG based on modulation instability was pushed down to 260 nm by making HCPCF with ultra-thin core-wall thicknesses of approximately 90 nm \cite{Gerome20}. Further details, including insights into resonant dispersive wave emissions spanning from the vacuum and deep UV regions, are available in \cite{Gerome19} with complementary studies based on hollow capillary fibers \cite{Gerome52}.

A striking illustration of the enabling power of HCPCFs came also from the demonstration of SCG in ambient air \cite{Benabid11} (see \cref{fig:HCPCF_SC}b). Although air has a nonlinear refractive index nearly three orders of magnitude smaller than glass, when confined in Kagome fibers it yielded multi-octave SC spectra under sub-millijoule pumping. This establishes HCPCFs as universal NLO platforms where essentially any gas can be harnessed for broadband frequency generation.

\subsection{Ultrashort pulse compression}
The ability of HCPCFs to deliver ultrashort, high-energy pulses has made them central to efforts in temporal pulse compression. Gas-filled HCPCFs can combine anomalous dispersion with nonlinear self-phase modulation to enable soliton self-compression down to few-cycle or even sub-cycle durations.

\begin{figure}[H]
\begin{center}
\includegraphics[width=1\columnwidth]{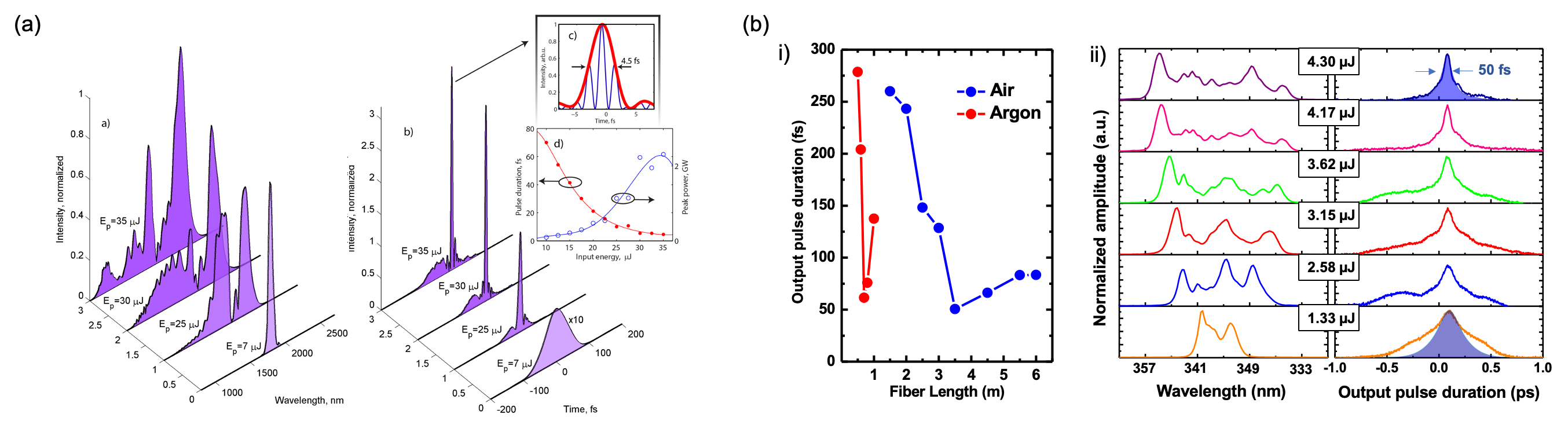}
\caption{(a) Sub-cycle compression experiments (data from \cite{Benabid13}) : Experimentally measured spectra (lhs) and output pulse profiles (rhs) for 0.2 m-long fiber filled with 4 bar of xenon.(b) UV pulse compression experiments (data from \cite{Gerome17}): i) Measured pulse-width with fiber length for air-filled  (blue curve) and 15 bar Ar-filled HCPCF (red curve), ii) Details of the output spectrum (lhs) and temporal trace (rhs) evolution with laser input energy for the case of 0.7 m-long fiber filled with 15 bar of argon. 
}
\label{fig:HCPCF_compression}
\end{center}
\end{figure}

The landmark experiment by Balčiūnas et al. \cite{Benabid13} demonstrated direct single-cycle self-compression of 80\,fs pulses at 1.8\,$\mu$m down to $\sim$4.5 fs inside a Kagome HCPCF filled with xenon (see \cref{fig:HCPCF_compression}a). This result was remarkable because it required no external compressors: the nonlinear dynamics inside the fiber naturally reshaped the pulses to the single-cycle limit while preserving carrier-envelope phase. 

Other works have extended this approach to higher average powers \cite{Benabid14, Gerome14}, various gases including air  \cite{Gerome15,Gerome16},  unconventional operating wavelength ranges such as UV \cite{Gerome17,Gerome18} (see \cref{fig:HCPCF_compression}b) and more advanced HCPCF-based compression schemes \cite{Gerome12,Gerome13}, establishing HCPCFs as versatile pulse compressors for driving extreme nonlinear optics.

These results mark a turning point: pulse compression is no longer limited by bulk optics or grating compressors but can be realized in a single fiber segment, providing compact, stable drivers for attosecond science and high-field physics.

\subsection{High-harmonic generation}
\begin{figure}[H]
\begin{center}
\includegraphics[width=0.9\columnwidth]{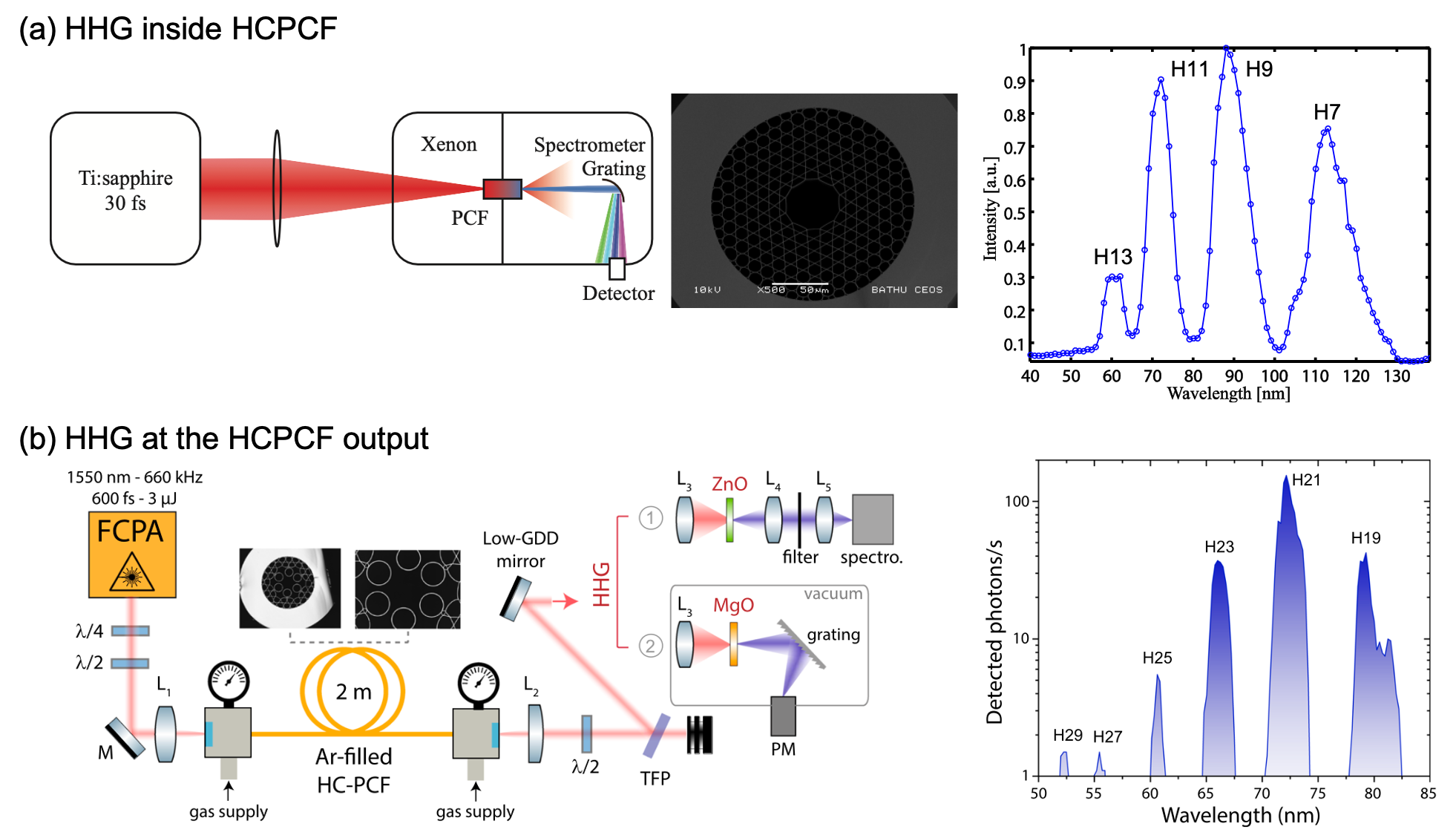}
\caption{(a) Generation of the 7th-13th harmonic of a Ti:sapphire laser system operating around 800 nm by propagating 30 fs pulses at $>$100 TW/cm$^2$ through a xenon-filled HCPCF (data from \cite{Benabid12}). (b) Generation of harmonics up to 29th of a sub-50 fs compressed pump laser operating around 1550 nm with intensity of 5.4 TW/cm$^2$ focalised at the argon-filled HCPCF output in a 200 $\mu$m thick MgO crystal (data from \cite{Gerome42}). 
}
\label{fig:HCPCF_HHG}
\end{center}
\end{figure}

HHG is a cornerstone of attosecond science, traditionally realized by focusing femtosecond pulses into free-space gas jets. The interaction length is limited by diffraction and by phase mismatch, and the required lasers are typically bulky, kilohertz-repetition-rate Ti:sapphire systems.
HCPCFs offer a paradigm shift by enabling HHG inside the waveguide. In 2009, Heckl et al. \cite{Benabid12} demonstrated the generation of harmonics up to the 13th order (wavelengths down to 50\,nm) in a xenon-filled Kagome fiber, with pulse energy thresholds as low as 200\,nJ (see \cref{fig:HCPCF_HHG}a). This reduced the required pulse energy by over an order of magnitude compared to free-space experiments. Importantly, HCPCFs also provide new phase-matching degrees of freedom: dispersion engineering, pressure tuning, or core-geometry modulation \cite{Benabid12}. Building on this, Gebhardt et al. demonstrated in 2021 that HHG during soliton self-compression could produce emission exceeding 300 eV, achieving an unprecedentedly high flux driven by a laser pump repetition rate of 98 kHz \cite{Gerome11}. In parallel, works showed that fiber-compressed pulses could directly drive HHG at the fiber output, enabling compact table-top extreme UV sources with high repetition rates. Such demonstrations have been achieved using ultrafast compressed laser pump operating at 1 $\mu$m \cite{Gerome43}, 1.55 $\mu$m \cite{Gerome42} (see \cref{fig:HCPCF_HHG}b) and 1.8 $\mu$m \cite{Gerome44} to respectively generate harmonics up to H25 (42 nm), H29 (52 nm) and H73 (25 nm).

These approaches hold promise for significantly improving conversion efficiency, potentially enabling fiber-integrated coherent EUV and XUV sources for applications in spectroscopy, imaging, and lithography.

\subsection{Plasma generation in gas-filled HCPCF}
Plasma, the fourth state of matter, consists of an ionized gas where free electrons and ions coexist and collectively respond to electromagnetic fields. Unlike solids, liquids, or neutral gases, plasmas exhibit rich electromagnetic behavior arising from their high charge mobility, space-charge separation, and long-range Coulomb interactions. This unique combination makes plasma not only a cornerstone of astrophysical and laboratory physics but also a highly nonlinear optical medium whose properties can be dynamically tuned by light itself. In this context, HCPCFs offer an exceptional microstructured platform for plasma generation and study. Their core, inherently filled with gas rather than solid glass, provides a natural environment where light and matter interact over long distances under strong confinement. The photonic microstructure confines the optical field within a micron-scale channel while isolating it from the silica cladding, thereby enabling stable plasma formation inside a guided geometry without damaging the fiber walls. 

Plasma generation inside gas-filled HCPCFs thus opens a distinct pathway to extreme-field nonlinear optics, bridging the regimes of ultrafast photoionization and sustained radio-frequency (RF) or microwave discharges. The tight transverse confinement and extended interaction length allow low-power ignition, stable confinement, and precise spatio-temporal control of electron density within the guided core. From a wave-propagation perspective, the free-electron population modifies the effective refractive index according to a Drude-type response, $n_{\textrm{eff}}\simeq\sqrt{1-\omega_p^2/\omega^2}$, $\omega_p\propto\sqrt{n_e}$ is the plasma frequency governed by the electron density $n_e$. In HCPCFs, this plasma–field overlap can be engineered through gas species, pressure, and excitation schemes. Plasma formation can occur either on ultrafast timescales (few-femtosecond photoionization) or as steady discharges driven by continuous RF or microwave fields—both realized within the same fiber-based architecture.

Over the past decade, two complementary routes have emerged for generating plasma inside HCPCFs: (1) Microwave-driven discharges, where surface-wave or resonator-assisted excitation at gigahertz frequencies sustains a quasi-steady plasma column along the hollow core; and (2) Ultrafast photoionization, where femtosecond or picosecond laser pulses induce transient plasmas that co-evolve with optical solitons in the fiber. These two regimes represent the twin pillars of plasma photonics in HCPCFs. Together, they have transformed the field by uniting gas discharge physics with guided-wave optics, enabling phenomena ranging from continuous-wave plasma emission in the UV–DUV to attosecond-scale light–matter interactions driven by single-cycle laser pulses.

\subsubsection{Microwave-driven plasma generation in HCPCF}
\begin{figure}[H]
\begin{center}
\includegraphics[width=0.9\columnwidth]{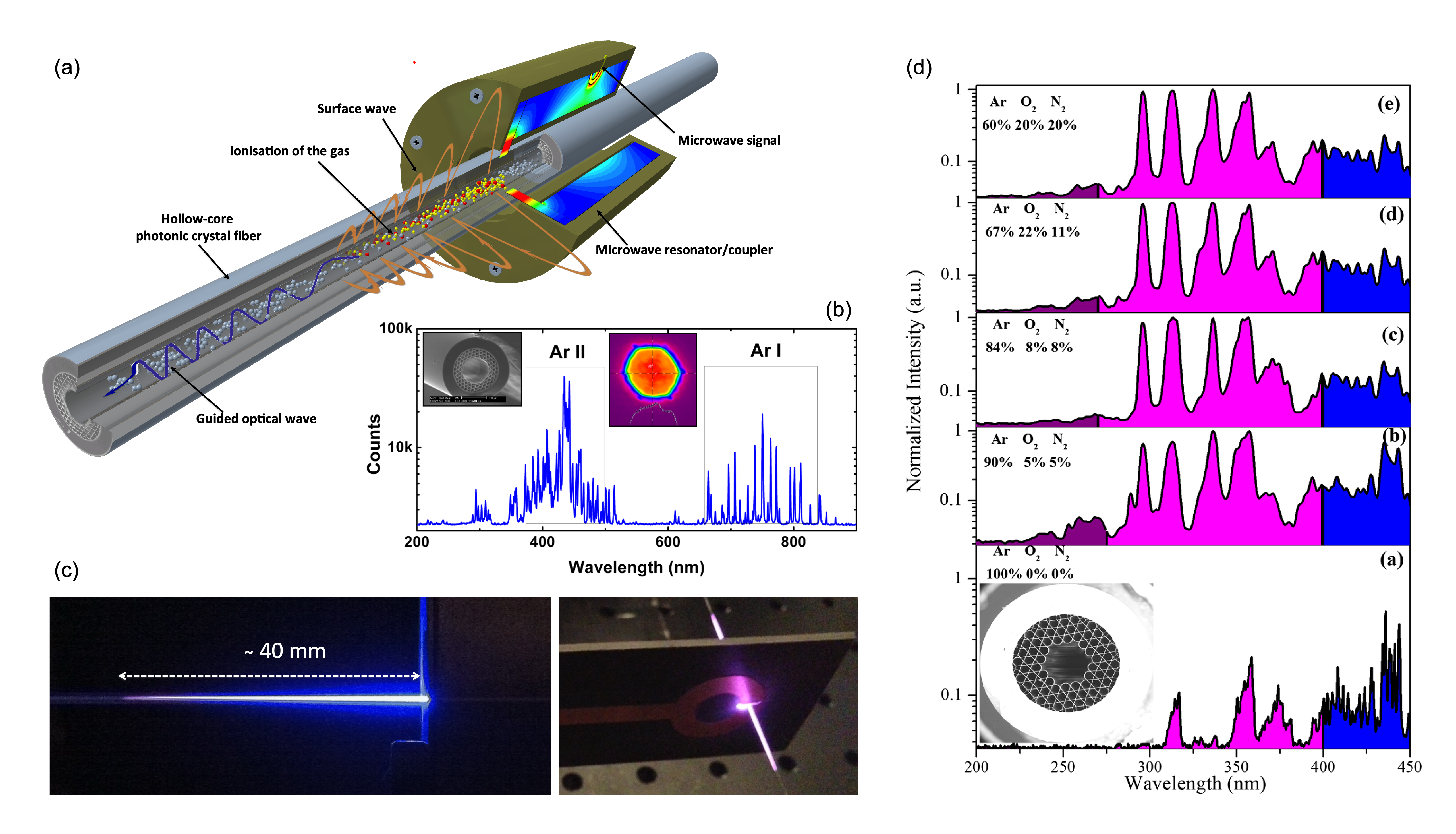}
\caption{(a) Schematic of plasma surface-wave excitation in HCPCF. (b) Spectrum of the emitted plasma light in a HCPCF (inset: cross section and output modal content of the plasma-filled HCPCF) (data from \cite{Benabid19}). (c) Pictures of plasma columns generated via Surfatron (left) and planar microstripline (right) excitators. (d) Evolution of the DUV/UV emission line spectra of an Ar-O$_2$-N$_2$ plasma created in a 19-cell Kagome fiber (SEM in inset) for different ratios of the gas mixture components (data from \cite{Benabid21}). 
}
\label{fig:HCPCF_plasma}
\end{center}
\end{figure}

The generation of plasma inside HCPCFs by microwave excitation represents a remarkable achievement in plasma photonics—combining high ionization rates with surprisingly moderate temperatures and full preservation of the silica structure. In their pioneering work, Debord et al. \cite{Benabid19} demonstrated the first confinement of a self-sustained microwave plasma column within an argon-filled Kagome HCPCF by coupling 2.45\,GHz power through a surfatron cavity (see \cref{fig:HCPCF_plasma}(a,b)). A surface-wave plasma propagated stably along several centimeters of fiber (\cref{fig:HCPCF_plasma}c), producing electron densities approaching 10$^{14}$\,cm$^{-3}$, while the gas temperature remained near 1300 K—well below the expected thermal load for such ionization levels. Intriguingly, the silica microstructure showed no damage, despite the plasma core temperature exceeding the glass transformation temperature by several hundred Kelvin.

This paradox — high ionization at low temperature with no structural degradation — was explained by the unique plasma confinement geometry inherent to the HCPCF. Because the plasma is localized in the center of the hollow core, it forms a thick, quasi-cylindrical space-charge sheath that isolates the hot plasma zone from the glass wall. The ions, having a mean free path comparable to or larger than the core diameter, transfer limited kinetic energy to the gas, which in turn limits wall heating. As a result, the core temperature remains moderate while electron density remains high, a condition rarely achievable in conventional plasma discharges.

Building on this foundation, Vial et al. \cite{Benabid20} introduced a split-ring resonator (SRR) launcher based on a planar microstripline, enabling efficient excitation of surface waves at the core boundary with only a few watts of microwave power. The SRR-based configuration yielded centimeter-scale plasma columns in various gases, reducing ignition thresholds and improving spatial control (\cref{fig:HCPCF_plasma}c). Later, Amrani et al. \cite{Benabid21} extended the concept to ternary gas mixtures (Ar/O$_2$/N$_2$), obtaining strong deep UV (200–450\,nm) plasma emission guided along the fiber (\cref{fig:HCPCF_plasma}d). Through kinetic modeling, they identified optimal mixtures (Ar $\sim$ 90\%, O$_2$ $\sim$ 5\%, N$_2$ $\sim$ 5\%) that enhance excitation efficiency while maintaining plasma confinement and stability.

Finally, the first observation of optical gain in a HCPCF was reported by Bateman et al. using DC excitation \cite{Gerome7}. More recently, in 2022, Gladyshev et al. introduced a novel, non-invasive microwave ignition approach by embedding a HCPCF within a slot in the side wall of a microwave waveguide \cite{Gerome8}. This innovation led to the first demonstration of a microwave gas-discharge fiber laser. He-Ar-Xe mixture was first used as the active gas medium \cite{Gerome9}, with a similar demonstration using He-Xe gas reported the following year \cite{Gerome10}.

Collectively, these works established microwave-driven plasma generation in HCPCFs as a singular physical regime—where plasma physics, photonic confinement, and materials science converge. The coexistence of high electron density, low gas temperature, and structural integrity stems directly from the self-organized sheath geometry of the plasma within the hollow core. This mechanism offers a new paradigm for creating stable, reconfigurable, and spectrally tunable plasma channels in photonic fibers, opening paths toward compact UV sources, plasma-assisted devices, and fundamental studies of strongly confined discharges.

\subsubsection{Ultrafast photoionization-induced plasma in HCPCF}
\begin{figure}[H]
\begin{center}
\includegraphics[width=0.9\columnwidth]{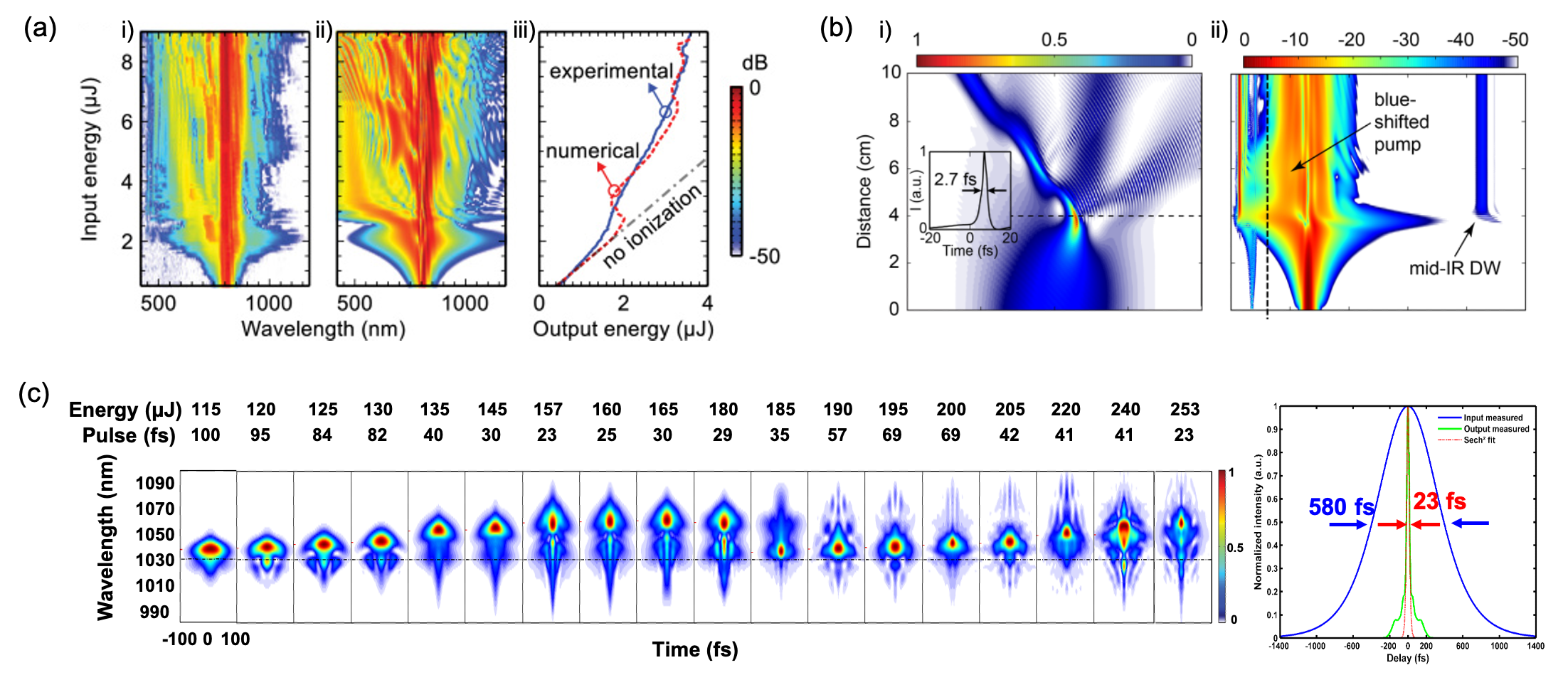}
\caption{(a) Results from \cite{Benabid9}: i) experimental and ii) numerical output spectra from a Ar-filled Kagome HCPCF as a function of input pulse energy and iii) comparison of the transmission with and without the ionization terms included. (b) Results from \cite{Benabid27} : i) temporal and ii) spectral evolution of a 30 fs pulse duration propagating in a Kr-filled Kagome HCPCF demonstrating a significant free-electron density role in the emission of mid-IR resonant radiation. (c) Spectral and temporal evolution of the output pulse with energy in an air-filled Kagome HCPCF. Autocorrelation traces of the compressed and input pulses recorded at 157 $\mu$J input energy (data from \cite{Benabid34}). 
}
\label{fig:HCPCF_bouncing}
\end{center}
\end{figure}

In addition to microwave-driven discharges, ultrafast photoionization in gas-filled HCPCFs provides a powerful means of generating plasma inside the fiber core. When high-energy femtosecond pulses self-compress to intensities near the ionization threshold, free electrons are created that dynamically modify the effective refractive index. This Drude-like contribution introduces a negative index shift that causes solitons to accelerate and undergo a persistent self-frequency blue-shift, in direct contrast with the Raman red-shift seen in molecular gases.
Seminal demonstrations in 2011 by Saleh et al. \cite{Benabid23} provided the theoretical foundation for this regime, predicting continuous soliton blue-shift in ionizing noble gases and mapping its competition with Raman effects. Shortly after, Hölzer et al. \cite{Benabid9} gave the first experimental evidence, observing blue-shifted soliton sequences in Ar-filled Kagome HCPCF (\cref{fig:HCPCF_bouncing}a). This established photoionization plasma as a distinct nonlinear actor in fiber optics. Numerical studies quantified the ionization threshold and its impact on soliton propagation \cite{Benabid10}.
Subsequent work highlighted how plasma can seed new instabilities and phase-matching regimes. Saleh et al. \cite{Benabid26} reported plasma-mediated modulational instability and asymmetric self-phase modulation. Novoa et al. \cite{Benabid27} and Ermolov et al. \cite{Gerome23} showed that plasma enables emission of mid-IR dispersive waves (\cref{fig:HCPCF_bouncing}b) and vacuum-UV supercontinua, respectively. Then, Köttig et al. \cite{Benabid29,Benabid30} demonstrated coherent plasma-induced soliton fission and petahertz-wide interference fringes. More recently, Suresh et al. \cite{Benabid31} introduced a pump–probe scheme that directly measured plasma electron densities along the fiber, linking them to spectral signatures. Further insights can be found in the recent review by John C. Travers \cite{Gerome19}.

Maurel et al. \cite{Benabid34} extended this paradigm to air-filled Kagome HCPCF, using 580\,fs, $\sim$100–250 $\mu$J pulses from a Yb laser at 1030\,nm. They observed a novel sequence of alternating red- and blue-shifts (\cref{fig:HCPCF_bouncing}c): at $\sim$115\,$\mu$J the pulse compressed and red-shifted to $\sim$100\,fs; at $\sim$157\,$\mu$J, further red-shift yielded stable self-compression down to 23\,fs. Increasing energy to $\sim$185 $\mu$J triggered a sudden plasma-induced blue-shift, followed by renewed compression, ultimately producing stable $\sim$23 fs pulses at 253\,$\mu$J. Remarkably, despite the plasma sensitivity, the compressed pulses were stable over 5 hours with only $\pm$0.3\,fs variation. This work showed that plasma generation can be harnessed, not avoided, to stabilize soliton self-compression in a single fiber stage. 

In follow-up work, the same team combined experiments with detailed numerical modeling using the multimode nonlinear Schrödinger equation (MMNLSE), including Kerr, Raman, and plasma terms \cite{Benabid33}. The simulations confirmed the observed “spectral bouncing” between red- and blue-shifts, attributing it to the interplay of Raman scattering, plasma formation, and excitation of higher-order modes. The numerical electron density maps revealed oscillatory plasma dynamics synchronized with soliton fission, consistent with the concept of floating solitons. The agreement between experiment and simulation firmly established plasma–Raman interplay as the governing mechanism behind stable compression to $\sim$20\,fs in air-filled fibers.
Together, these studies demonstrate that plasma formation inside HCPCFs is not merely a loss channel but a stabilizing and tunable mechanism for pulse compression. The controlled interplay between Raman red-shift and plasma blue-shift enables robust few-cycle pulse generation from simple Yb-based systems, positioning photoionization plasmas as a key element in the future of fiber-integrated attosecond drivers and plasma-assisted dispersion engineering.

\subsection{Non-classical light generation}

Beyond their well-established role in classical nonlinear optics that we covered in previous sections, such as enabling soliton dynamics, self-phase modulation, and plasma formation, gas-filled HCPCFs have recently emerged as exceptional platforms for the generation and manipulation of non-classical light. Their geometry, which confines light within a gas-filled microstructured core, provides a long interaction length, high field confinement, and freedom from solid-state noise, offering conditions where quantum coherence and strong light-matter coupling can develop in entirely new ways. 

\begin{figure}[H]
\begin{center}
\includegraphics[width=0.9\columnwidth]{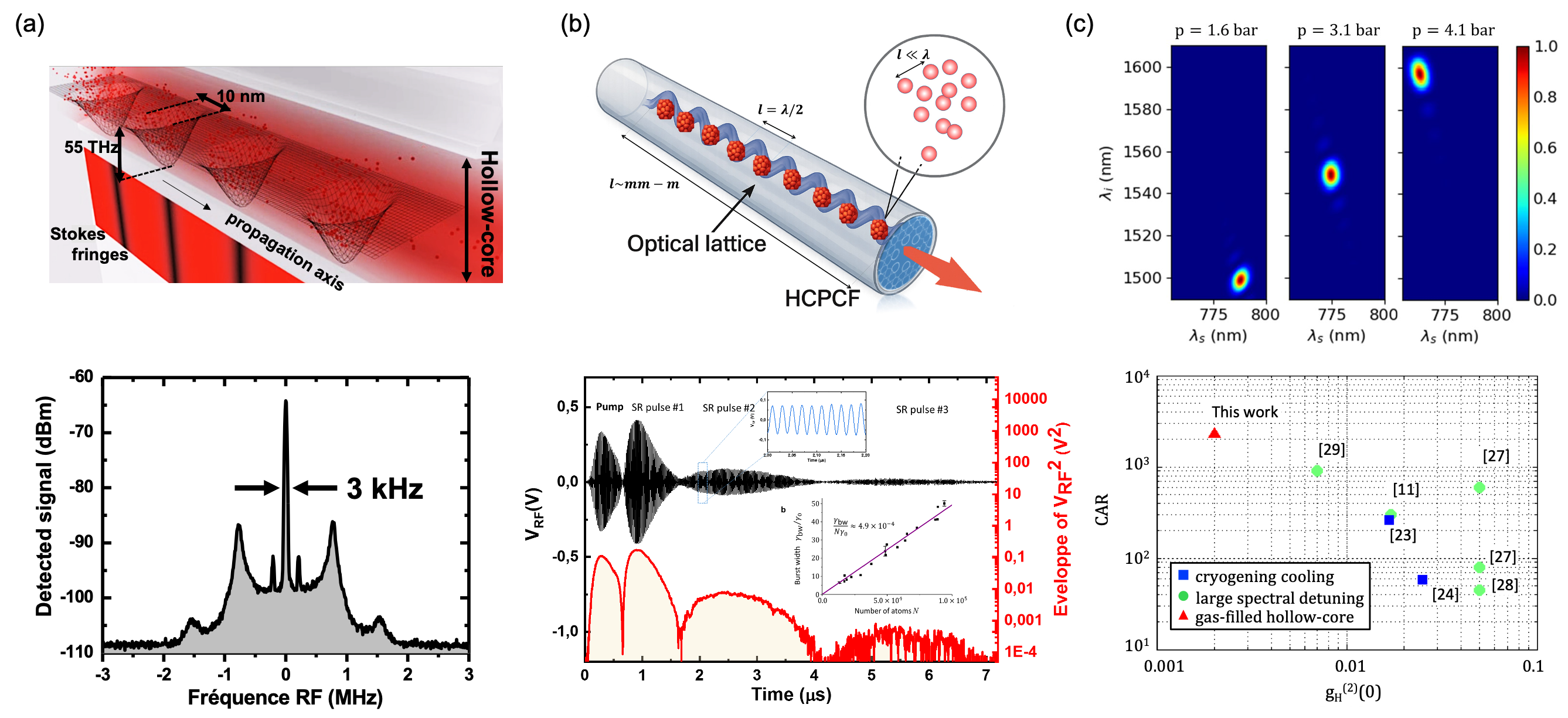}
\caption{(a) Illustration of the Raman gas self-organizing into deep nano-trap lattice concept (see the video). Spectra showing a linewidth of the transmitted forward Stokes for a CW pump power of 12 W and gas pressure of 20 bar 30 m-long hydrogen-filled HCPCF as low as $\sim$3 kHz, more than five orders of
magnitude narrower than conventional-Raman pressure-broadened linewidth (data from \cite{Benabid15,chafer2017macroscopic}). (b) Schematic illustration of ultracold ${}^{88}$\textit{Sr} atoms loaded into an optical lattice and then transported inside the HCPCF. The superradiant emission is enhanced by the multimode optical guidance of the fiber (data from \cite{okaba2019superradiance}). (c) Tuneability of the simulated joint spectral amplitude in a xenon-filled HCPCF via pressure control, and the corresponding performance of the Raman-free fiber-based photon-pair source compared to the state-of-the-art (data from \cite{Benabid32}, the references indicated in the plot are included there). 
}
\label{fig:HCPCF_NewLight}
\end{center}
\end{figure}

A first glimpse of this quantum regime was provided by \cite{Benabid15}, who showed that continuous-wave stimulated Raman scattering in hydrogen-filled photonic-bandgap HCPCFs can drive the gas into a self-organized, nanostructured state. Interference between forward and backward Stokes waves creates a stationary Raman coherence grating forming a deep optical lattice of $\sim$15 THz potential depth and nanometric periodicity, within which molecular hydrogen becomes localized in tens of thousands of nanoscale traps. The trapped molecules scatter in the Lamb-Dicke regime, producing an ultranarrow Stokes line ($\sim$14 kHz, a linewidth reduced to $\sim$3 kHz by Chafer et al. \cite{chafer2017macroscopic}) with Rabi splitting, translational sidebands, and four-wave-mixing satellites-clear signatures of quantum motion in a strongly coupled light-matter ensemble (\cref{fig:HCPCF_NewLight}a). This experiment represented the first observation of the Dicke regime in a dense gas at room temperature, showing that a HCPCF can act as a hybrid quantum medium where molecular and photonic degrees of freedom become collectively organized.

A conceptually related manifestation of collective emission was later achieved by  Okaba et al. \cite{okaba2019superradiance}, who demonstrated superradiance from ultracold ${}^{88}$\textit{Sr} atoms trapped inside a Kagome-lattice HCPCF (\cref{fig:HCPCF_NewLight}b). By exciting the narrow ${}^1\!S_0-{}^3\!P_1$ intercombination line, they observed emission bursts up to fifty-five times faster than single-atom decay, confirming cooperative light emission. Remarkably, the atomic ensemble spontaneously organized its radiation field into a specific superposition of guided fiber modes that maximized coupling to the atoms, leading to an effective cooperativity near unity—an emergent self-mode-matching effect unique to guided-wave superradiance. 

Moving from collective emission to engineered quantum states, Cordier et al. \cite{cordier2019active} demonstrated photon-pair generation and heralded single-photon emission in noble-gas-filled IC HCPCFs. By tuning gas pressure and exploiting the fiber’s multiband dispersion, they could tailor the joint spectral intensity of photon pairs from entangled to factorable states, and in Xenon achieved record-high purity and a coincidence-to-accidentals ratio exceeding 2700, with $g^2(0)=0.002$ \cite{Benabid32} (\cref{fig:HCPCF_NewLight}c). Because the optical mode interacts almost exclusively with the gas, these sources are intrinsically Raman-free, stable, and spectrally tunable across visible and telecom bands. Later, Tyumenev et al. \cite{tyumenev2022tunable} extended this paradigm to quantum frequency conversion, using hydrogen-filled HCPCFs to shift single-photon frequencies by $\sim$125 THz via Raman coherence while preserving their quantum correlations with near-unity efficiency. Finally, Finger et al. \cite{Benabid16} demonstrated the generation of bright twin-beam squeezed vacuum in argon-filled Kagome fibers, achieving record brightness levels thanks to the absence of Raman noise. Together, these advances trace a coherent trajectory from molecular self-organization and atomic superradiance to the generation and processing of individual photons, revealing the extraordinary versatility of gas-filled HCPCFs as quantum photonic media. They establish the foundation of a new discipline—quantum gas photonics—in which light and matter co-evolve within a fiber to produce, shape, and translate quantum states with unprecedented control and purity.

\section{Conclusions \& Perspectives}
\label{sec:conclusion}

In conclusion, in this work we reviewed the key physical mechanisms and recent advances in extreme nonlinear optics within multimode and hollow-core fiber platforms. By elucidating the interplay of plasma formation, filamentation, and spatiotemporal wave dynamics under intense femtosecond excitation, we specifically highlighted how these effects underpin ultrabroadband supercontinuum and high-harmonic generation. The discussed developments point toward new opportunities for ultrafast photonics, quantum light sources, and next-generation optical communication technologies. Finally, we outline in this section our perspectives and the emerging directions, which are expected to drive further breakthroughs in extreme nonlinear fiber optics in the near future.

\subsection{Perspectives: multiphoton ionization effects in multimode fibers}

The investigation of MPI in multimode optical fibers is likely to open new directions that go well beyond the observation of plasma effects alone. In the last few years, femtosecond laser exposure has proven to be not only a diagnostic tool for studying nonlinear light–matter interaction in glass, but also a means for engineering the material itself. The possibility of inducing local and permanent refractive index variations inside the fiber core suggests that laser micromachining could become an intrinsic part of the fiber fabrication process, rather than a post-processing step. In this sense, one can imagine the possibility of directly writing micro- or even nano-structured regions within the fiber volume, defining new functionalities without removing material.
An equally promising aspect concerns the controlled generation of plasma filaments. When the balance between Kerr self-focusing and plasma defocusing is properly achieved, the resulting self-channeling regime can sustain light propagation along narrow paths that may be straight or helical, depending on the launch conditions. The observation of helical plasma filaments, obtained by tilting the input beam or imposing an offset, indicates that it might be possible to reproduce and stabilize such structures in a predictable way. These helically modified regions could later act as guides for secondary optical modes or even as templates for twisted photonic cores.

From a broader perspective, these studies also touch on the field of optical manipulation. The same nonlinear propagation regimes that give rise to plasma filaments and supercontinuum emission can, under suitable conditions, produce spiral or ring-shaped beams carrying OAM. Their structured intensity and phase profiles naturally lend themselves to optical trapping applications. In particular, the coexistence of intense localized fields and wavelength-dependent emission suggests that fiber-based optical tweezers may be realized without the need for bulky free-space optics.
In the near future, progress will likely come from combining femtosecond micromachining and optical trapping concepts within a single experimental framework. Understanding the cumulative effects of repeated MPI exposure, the role of thermal relaxation, and the limits of damage control will be essential for translating these laboratory observations into practical photonic components. Eventually, multimode fibers may evolve into multifunctional platforms where light is able to modify, sense, and manipulate matter within the same physical structure.

Finally, studying the MPA-induced UL provides a powerful route for assessing the optical quality of glassy and photonic materials. Because UL originates from the interaction of intense light with intrinsic or processing-induced defects, its spectral features and spatial distribution directly reflect the presence, type, and density of such imperfections. By analyzing these luminescent signatures—such as emission intensity, wavelength shifts, and pattern periodicity—it becomes possible to correlate the optical response with structural integrity and compositional homogeneity. This approach offers a non-destructive means to characterize materials and fibers, guiding optimization of fabrication processes and enabling the development of components with enhanced transparency, stability, and nonlinear performance. In this sense, the investigation of UL not only deepens the understanding of defect-related phenomena, but it also opens new perspectives for quality control and design in advanced photonic systems.

\subsection{Perspectives: azimuthal instabilities, breather and light bullets}

\begin{figure}[H]
\begin{center}
\includegraphics[width=0.7\columnwidth]{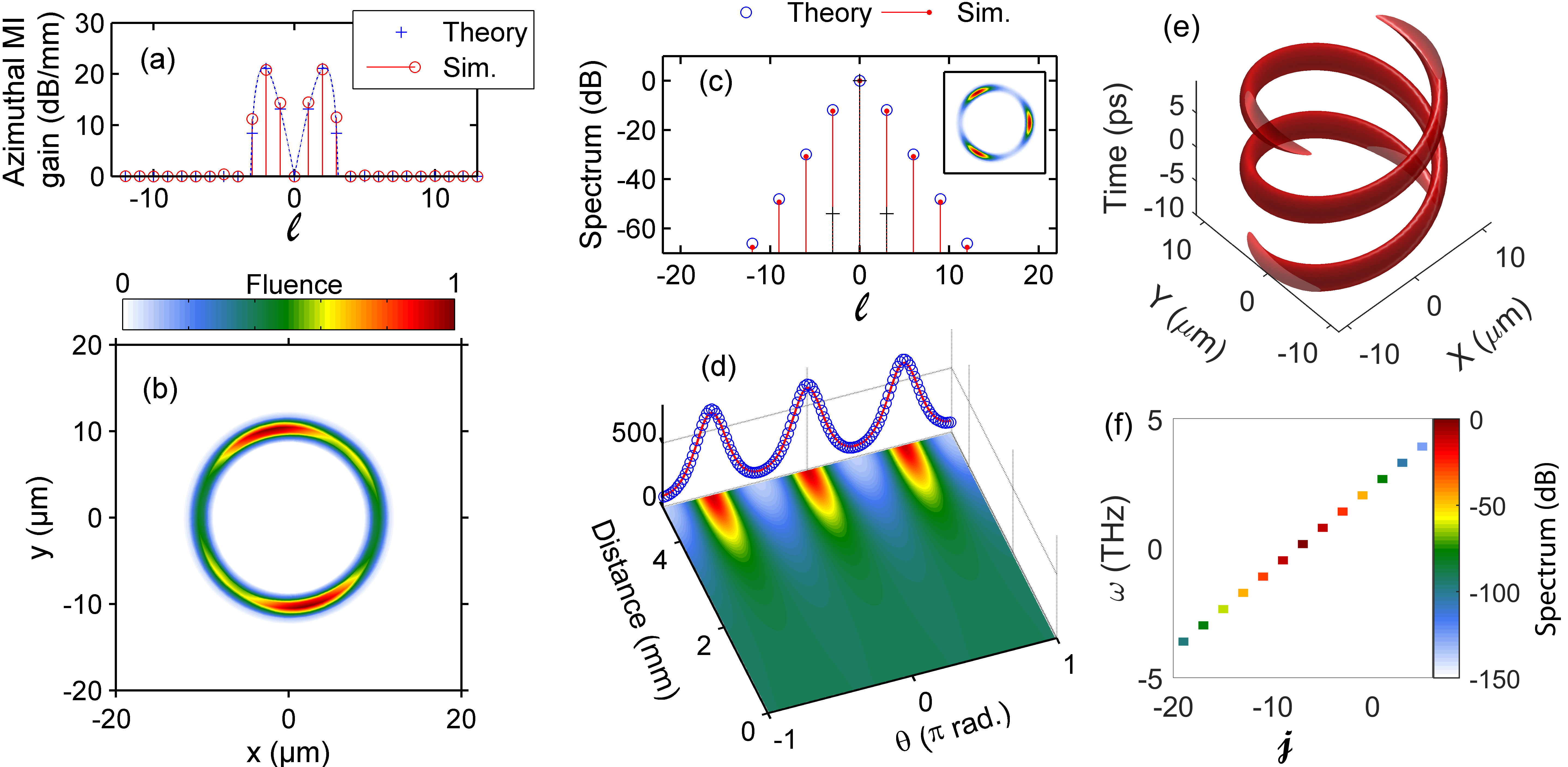}
\caption{(a-b) Spontaneous development of azimuthal modulation instability. (a) Example of azimuthal MI gain calculated theoretically and numerically in a ring-core fiber for an input power of 300 kW. (b) Corresponding output normalized fluence distribution in space after MI development. (c-d) Emergence of Akhmediev breathers in the regime of seeded azimuthal modulation instability. (c) Input and output power spectrum profiles (inset: output normalized fluence). (d) Evolution of spatial power profile along propagation distance (seeding at $\ell=\pm$3), and corresponding theoretical and simulated profiles at maximum breather compression. (e-f) Iso-contour at half-maximum of a spacetime Jacobi-dn light bullet exhibiting a carrier orbital angular momentum $\ell_0=-6$ and its associated mode-resolved spectrum. In panel (f), the modes are classified by their respective total angular momentum $j$. 
}
\label{fig:STSoliton}
\end{center}
\end{figure}

As the input power approaches the critical threshold, intermodal coupling effects become increasingly pronounced. For instance, recent studies have demonstrated that azimuthal spontaneous modulation instability can induce substantial energy exchange between orbital angular momentum modes in ring-core fibers \cite{MIOAMKibler} [see also Fig. \ref{fig:STSoliton}(a,b)]. When seeded, the azimuthal modulation instability can also lead to the generation of Akhmediev breathers [see Fig. \ref{fig:STSoliton}(c,d)]. Remarkably, this phenomenon closely parallels the modulation instability observed in single-mode fibers operating in the anomalous dispersion regime. Specifically, it has been shown that in ring-core fibers, the nonlinear propagation dynamics of the field’s angular distribution can be accurately modeled by a NLS equation. Because the propagation constant in such fibers is a concave function of the angular momentum $\ell$ with an almost constant negative curvature, the propagation of an angular wavepacket—namely, a field formed by a coherent superposition of $\ell$-modes—necessarily occurs within the anomalous angular dispersion regime. 

However, it is interesting to note that the periodic and finite nature of the angular dimension over which the dynamics takes place nevertheless implies some differences in behavior with restect to time-domain modulation instability. For instance, the fact that the orbital angular momentum spectrum is discrete by nature implies that the modulation instability process has a power threshold below which the effect vanishes. As recently discussed \cite{BejotVectoriel}, the discrete nature of the OAM spectrum is also responsible for the appearance of angular breathers without the use of a deterministic and coherent seeding, as needed for their time-domain analogues.

All these findings open new perspectives for photonic functionalities, such as all-fiber OAM converters and amplifiers. While these effects have been investigated primarily in the monochromatic regime, complex spatiotemporal dynamics are also expected to emerge in the strongly nonlinear regime. For example, recent studies have demonstrated four-wave-mixing-based parametric amplification of spin–orbit modes in ring-core fibers \cite{MI_OAM_Murphy}. This process, during which both energy and total angular momentum are conserved, can only be explained within a full vectorial framework, i.e., beyond the weak-guidance approximation \cite{BejotVectoriel}, a regime where spin-orbit couplings take place \cite{RamachandranSOC,KarolinaSOC}. This parametric process is inherently linked to space-time modulation instabilities that lead to the amplification of specific frequencies in specific angular momentum modes, whatever the (frequency) dispersion regime. As recently demonstrated \cite{BejotSOLITON_ST}, a direct corollary is the existence of spacetime solitons (i.e., light bullets) in ring-core fibers, whatever the dispersion regime [see Fig. \ref{fig:STSoliton}(e,f)].

\subsection{Perspectives in gas-filled HCPCFs}
As we have seen, the combination of ultra-low transmission loss, ultra-low optical overlap with the cladding glass and long interaction lengths has enabled gas-filled HCPCFs to evolve over the past 2 decades into a uniquely powerful platform for extreme nonlinear and quantum optics. They enable laser intensities and energies far beyond the limits of solid materials, support broadband light–matter interactions from the vacuum ultraviolet to the mid-infrared, and host gas-phase media ranging from ultracold atoms to hot plasmas with unprecedented spatial and temporal control. Despite these remarkable achievements and growing maturity, HCPCFs continue to offer fertile ground for new science and technology, and potentially revolutionary prospects in relativistic photonics, ultrafast spectroscopy, exotic nonlinear dynamics, quantum optics, and emergent many-body phenomena.\\
HCPCFs continue to push light source spectral coverage and bandwidth limits. Already, mid-IR pumping in noble-gas HCPCFs has produced coherent spectra spanning from $\sim$350\,nm to 5\,$\mu$m \cite{Gerome21}, covering most of the DUV-mid-IR gap. Future fiber designs (e.g., with optimal glass surface roughness \cite{osorio2023hollow} in IC HCPCF) and new pumping schemes could extend this further into the extreme UV ($<$100\,nm) and far infrared ($>$10\,$\mu$m). Such developments will yield compact, fiber-based sources for molecular spectroscopy and imaging across electronic and vibrational bands.
Gas-filled HCPCFs open access to extreme regimes of nonlinear optics. Noble-gas HCPCFs have already enabled tabletop HHG getting closed to the soft X-ray range. By optimizing dispersion, pressure profiles and plasma absorption, future experiments may generate even higher harmonics or attosecond pulse trains inside fibers. At the same time, HCPCFs are being used as plasma waveguides. The first demonstration by Debord et al. \cite{Benabid19} of microwave-plasma generation and micro-confinement within a large-core ($\sim100$ $\mu$m diameter) HCPCF could transform this "plasma-core" fiber into a reconfigurable waveguides for intense beams or enable novel plasma-assisted effects, effectively uniting fiber optics with plasma photonics.
Also, the extraordinary damage threshold of HCPCFs means they can guide pulses at near-relativistic intensities. Modeling shows that a few-joule, few-femtosecond pulse in a 100 $\mu$m diameter core could reach the onset of relativistic optics ($\sim10^{18} W/cm^{2}$). In the future, specially designed large-core fibers and optimized coupling could allow experiments with petawatt-level guided pulses, opening prospects for fiber-based wakefield acceleration or strong-field QED in a stable waveguide. In short, gas fibers may become testbeds for \textit{relativistic photonics}, enabling phenomena like laser-driven particle beams or novel X-ray sources under well-controlled conditions. 

\begin{figure}[H]
\begin{center}
\includegraphics[width=1\columnwidth]{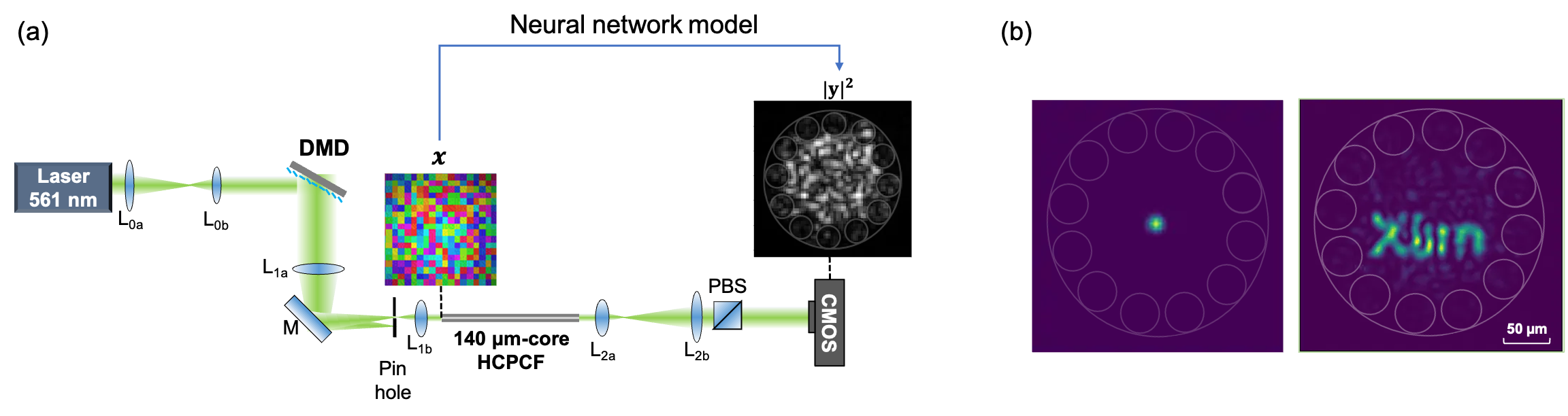}
\caption{(a) Experimental setup developed for machine learning of a multimodal HCPCF model using a neural network model. (b) Examples of optimized output distributions from a 140 $\mu$m hollow-core fiber using the hologram of a spatial light modulator: on the left, a 10 $\mu$m-diameter focal spot, and on the right, a projection of the Xlim logo (data from \cite{Gerome41}). 
}
\label{fig:HCPCF_multimode}
\end{center}
\end{figure}

Multimode guidance in gas-filled HCPCFs represents one of the most transformative directions for the future of light–matter science. Unlike traditional fibers, the inherent gapless modal structure of HCPCFs naturally supports higher-order modes, allowing for structured light to propagate over long distances thanks to it low dispersion values with minimal loss. Far from a limitation, this multimode nature—especially in enlarged-core designs—opens a new paradigm: not just for spatial multiplexing or high-energy pulse delivery, but for a deeper convergence between advanced light shaping and the structured response of gas-phase media. Recent demonstrations of neural network-based wavefront control and spatial light modulator induced modal synthesis confirm that today's photonic technologies—ranging from spatial light modulator class beam conditioning to AI-assisted adaptive optics—are now capable of harnessing the full potential of HCPCF modal landscapes as shown in \cref{fig:HCPCF_multimode} \cite{Gerome41}. In this context, the self-organized phenomena already observed—such as spatio-temporal intermodal dynamics \cite{Gerome28}, self-cleaning \cite{Gerome29}, superradiant atomic bursts \cite{okaba2019superradiance}, Raman-induced nanotraps and self-assembled molecular lattices \cite{Benabid15}—may be only the earliest glimpses of a much broader class of emergent and self-organised phenomena. With precise control over both spatial, spectral and temporal degrees of freedom, and with gases serving as tunable nonlinear and quantum materials, HCPCFs are poised to become platforms where structured light and structured atoms co-evolve. We anticipate that the fusion of multimode guidance, gas-phase engineering, and emerging machine-intelligence-driven control will usher in a new scientific territory—where nonlinear optics, quantum science, and many-body physics converge in a fully fiber-integrated format.

\vspace{1cm}
Acknowledgment : FG, BD, FB thank their colleagues from GPPMM group and GLOphotonics (Foued Amrani, Mathieu Chafer, Frédéric Delahaye  and Kostiantyn Vasko) for their help in data collections and processing. BK and PB thank Karol Tarnowski and Karolina Stefa\'nska for fruitful collaborations. 

This work was supported by the European Innovation Council (101185664), the Italian Ministerial grant PRIN2022 "SAFE" (2022ESAC3K), the European Union via the Marie Skłodowska-Curie Actions under the BESCLING project (grant No. 101209943), the EIPHI Graduate School (contract "ANR-17-EURE-0002"), and benefited from the facilities of the SMARTLIGHT platform funded by the Agence Nationale de la Recherche (EQUIPEX+ contract "ANR-21-ESRE-0040") and R\'egion Bourgogne Franche-Comt\'e and Région Nouvelle Aquitaine.





\bibliographystyle{unsrtnat}
\bibliography{biblio}

\end{document}